\documentclass[pdflatex,sn-mathphys-num]{sn-jnl}

\usepackage{graphicx}%
\usepackage{multirow}%
\usepackage{amsmath,amssymb,amsfonts}%
\usepackage{amsthm}%
\usepackage{mathrsfs}%
\usepackage{xcolor}%
\usepackage{geometry}
\usepackage{libertinus}  
\usepackage{libertinust1math}
\usepackage[T1]{fontenc}

\begin{document}

\title{Fine-grained dynamics of entanglement in non-integrable quenches far across the Ising quantum critical point}
\author*{Aditya Banerjee}\email{adityabphyiitk@gmail.com}
\affil{Theory Division, Saha Institute of Nuclear Physics, 1/AF Bidhannagar, Kolkata 700064, India}

\abstract{The task of exploring and understanding various aspects of far-from-equilibrium dynamics of closed and generic quantum many-body systems has received a thrust of attention in recent years, driven partly by remarkable advances in ultracold experimental technologies. In this work, for the paradigmatic Ising spin chain with transverse and longitudinal fields and partly motivated by the practice of site-resolved control in contemporary ultracold experiments, we present numerical observations of several $\textit{fine-grained}$ (small-subsystem level) features of far-from-equilibrium dynamics from a quantum informational point of view, induced by quantum quenches far across the Ising critical point between states deep inside the para- and ferro-magnetic regimes. Rather featureless dynamics is seen for ferromagnetic to paramagnetic quenches, but paramagnetic to ferromagnetic quenches exhibit rich behaviour, including recurrences of an approximately Page-like dynamics of entanglement entropies of one- and two-spin subsystems, periodic but short-lived occurrences of approximately $1-$uniform states, a series of sudden deaths and revivals of entanglement between two spins in the system's bulk, non-analytic cusps in single-copy entanglement entropy for three-spin and bigger subsystems, insufficient mixedness and a series of scrambling-$\textit{un}$scrambling of local mutual information between neighboring spins. Moreover, essentially indistinguishable dynamics is seen at very early times between the integrable limit (zero longitudinal field) and non-integrable cases, with the former eventually showing signatures of better mixing and faster approach to equilibration than the latter. These features are expected to hold for quench dynamics across Ising quantum critical points in more complicated systems.}


\maketitle

\section{Introduction} \label{sec1}

The search for prominent and prevalent phenomena, and a set of organizing principles and frameworks with which to understand them, in the regime of far-from-equilibrium dynamics of quantum many-body systems continues to be an exciting enterprise that has received an acceleration in the last decade or so, thanks largely to breakthroughs in ultracold experimental technologies, see e.g. \cite{Lewenstein2012book,Gross2017,Ueda2020,Schafer2020}. Even for the classes of quantum many-body systems whose equilibrium properties, such as their ground state structures and phase transitions between them, are well understood in substantial detail within unified and organizing frameworks \cite{Sachdev-QPTbook,DuttaQPTbook}, probing, predicting and explaining various aspects of their far-from-equilibrium dynamics is a substantially less mature field, especially in the non-integrable and non-critical regimes where appropriate and widely applicable analytical approaches are scarce. 
\vspace{0.2cm}

A widely applicable and general framework, that is agnostic to system-specific details in its formulation, is quantum information theory \cite{Wilde2017,Hayashi2017,Holevo2019}. It may be hoped that deploying various tools and notions from this framework to probe non-equilibrium dynamics would be illuminating and may reveal interesting new aspects, and eventually might prove an indispensable guide in the search for deeper principled understanding. Indeed, this has already been the case for non-equilibrium dynamics involving quantum critical systems (see the relevant mention with references in the next paragraph). Moreover, irrespective of the underlying physical systems and mechanisms, the dynamics of quantum information, manifested in non-equilibrium few- or many-body quantum systems, is in itself of great interest both theoretically and experimentally, and from both fundamental and technological perspectives, see e.g. \cite{Swan2019,Frerot2023}.
\vspace{0.2cm}

A common way to throw a quantum system far out of equilibrium is by suddenly quenching one of its parameters, and this is most interesting when the initial state is very far from the equilibrium ground state of the post-quench Hamiltonian \cite{Polkovnikov2011,Mitra2018}. Naturally the system will then try to attain equilibrium over the due course of time and several questions about the characteristics and mechanisms of its non-equilibrium dynamics may be asked and investigated. Much is understood at this time when the Hamiltonian driving the dynamics is integrable \cite{Essler2016,Alba2018,Wu2021,Bertini2022} and/or critical \cite{Calabrese2005,Polkovnikov2011,Calabrese2016}, due significantly to the availability of widely applicable analytical methods to deal with these regimes. In particular, main signatures in the latter scenario include efficient scrambling of local quantum information and operator-spreading, and near-ballistic growth of bipartite entanglement until saturation and light-cone spreading of correlations, see e.g.\cite{Calabrese2005,Chiara2006,Lauchli2008,Manmana2009,Gring2012,Cheneau2012,Langen2013,Kim2013,Jurcevic2014,Bonnes2014,Kaufman2016,Polkovnikov2011,Mitra2018}, and a picture based on ballistically propagating quasiparticles is understood as the generic underlying mechanism for this regime \cite{Calabrese2005,Rieger2011,Coser2014,Calabrese2016}. Rigorous and general upper bounds for the growth of bipartite entanglement entropies for generic Hamiltonians with and without symmetries have also been established, see e.g. \cite{Bravyi2007,Acoleyen2013,Marien2016,Shi2024,Toniolo2025,Rakovszky2019,Huang2020,Znidaric2020}. Of course, quench dynamics of a large number of non-integrable systems falls outside the aforementioned paradigm (and the dynamical growth of their entanglement entropies are well below those upper bounds), owing to a variety of factors that are too many to mention here. One prominent factor that plays an underlying role in this article is the confinement of excitations, of which the mixed field Ising spin chain is the simplest and a prime example \cite{McCoy1978,Delfino1996,Fonseca2003,Rutkevich2008}.
\vspace{0.2cm}

In this work, we numerically study aspects of paramagnetic-to-ferromagnetic quenching of the Ising spin chain, and will compare this case to the case of the opposite quench. Comparisons between the integrable and non-integrable cases will also be presented. The quenches we study in this work are between parameter regimes that reside deep inside the paramagnetic and ferromagnetic phases, i.e., far across the Ising critical point, and hence the considered quench protocols are very non-perturbative, and can thus be considered as "intense" quenches. A significant focus is on the quantum informational behaviour of the dynamics of small subsystems, which is what we mean by "fine-grained" features in this work. Attributes of particular concern to us are a so-called single-copy entanglement entropy which depends only on the leading eigenvalue of a density matrix and is therefore related to the ground state of the so-called entanglement Hamiltonian, bipartite mixed state entanglement as measured by the negativity of entanglement between constituents of small subsystems comprising of a few spins, local information scrambling amongst nearby spins as measured by tripartite mutual information, and "mixedness" of small subsystems as quantified by the (satisfaction or violation of) majorization relations. Another sense in which we mean "fine-grained" features in this work is by our consideration of the dynamics of quantities involving a subset of the entanglement spectrum (eigenvalues of reduced density matrices) corresponding to the considered subsystems, i.e. single-copy entanglement entropy and majorization relations, which highlight the finer details of their dynamics through the behaviour of the leading members of the entanglement spectra. Dynamical behaviour of the entanglement spectra (or its leading members or of certain functions of them) and other quantum informational attributes of subsystems have previously revealed other characteristics of quench dynamics and the approach to equilibration and/or thermalization or lack thereof, see e.g. \cite{Vinayak2012,Chamon2014,Torlai2014,Canovi2014,Yang2017,Chang2019,Surace2020,Kvorning2022,Zhang2025}. 
\vspace{0.2cm}

Putting to use the notion that subsystems of any closed quantum many-body system are by definition open quantum systems interacting with environments which are the remainder of the whole system, and by employing elementary notions from the theory of open quantum systems, exemplified with the mixed field Ising spin chain, it was recently demonstrated in \cite{Banerjee2025} that the dynamics of sufficiently small subsystems in the case of paramagnetic-to-ferromagnetic quench was strongly non-Markovian (memory-retention and information backflows), while the opposite quench exhibited essentially Markovian (memory-less and monotonic information flow) subsystem dynamics. Moreover, with increasing level of non-integrability and/or increasing subsystem sizes, the dynamics of a certain function quantifying a measure of distance between subsystem density matrices' eigenvalues showed remarkably systematic behaviour. This work reveals entanglement-related additional subtle characteristics of the dynamics of subsystems of the mixed field Ising model.
\vspace{0.2cm}

We will see that paramagnetic-to-ferromagnetic quenches exhibit periodic occurrences of Page-like dynamics of entanglement entropies of one- and two-spin subsystems along with an almost-periodic albeit short-lived occurrences of (approximately) $1-$uniform states, mixed state bipartite entanglement between two spins (as measured with entanglement negativity) shows sudden deaths and subsequent revivals, tripartite mutual information shows strongly oscillatory behaviour indicating not just scrambling but also \textit{un}scrambling of local information occurring periodically, existence of non-analytic cusps in the single-copy entanglement entropy, and finally the lack of satisfaction of majorization relations across all times amongst the eigenvalues of subsystem density matrices, potentially providing another evidence of subsystems' dynamics being non-Markovian. Since all of these features show up only in paramagnetic-to-ferromagnetic quenches, we consider them to be features that (likely necessarily) accompany subsystem non-Markovianity demonstrated in \cite{Banerjee2025}. On the other hand, the opposite quench will be seen to be featureless with regards to these quantities. In addition, comparisons with the integrable case reveal that the dynamical differences in these attributes between integrable and non-integrable cases is negligible in the very early times, but subsequently the former shows better mixedness in the dynamics of subsystems and an apparently faster approach to eventual equilibration by exhibiting for e.g. an earlier escape from the persistently oscillatory behaviour of entanglement quantifiers within our simulation times.
\vspace{0.2cm}

Our method of choice for simulating the non-equilibrium quench dynamics is the second-order time-evolving block decimation (TEBD2) approach within the framework of matrix product states (MPS) \cite{Vidal2004,Schollwoeck2011,Paeckel2019,Xiang2023}. The advantages of the MPS methods are well known - they allow access to large system sizes (beyond the capacities of exact diagonalization or contemporary quantum algorithms) while being quasi-exact in precision, and are not afflicted by issues of Monte Carlo methods such as the complex-phase problem due to real-time evolution. The chief limitation of MPS methods come from their inability to quasi-exactly represent sufficiently highly entangled states \cite{Osborne2006,Eisert2006,Schuch2008,Schuch2008a}, however given the oscillatory dynamics of entanglement that will be seen in the paramagnetic-to-ferromagnetic quench, this limitation will not be of relevance to us in this work.
\vspace{0.2cm}

This work focuses mostly on the dynamics of small subsystems of a few spins. This is motivated by at least three factors. Firstly, as already mentioned previously, one of the aims of this study is to obtain a "fine-grained" picture of the dynamics (of quantum entanglement and information) at the localized level of a few spins. Secondly, in contemporary quantum informational and ultracold experiments, site-resolved control and measurement of quantum systems is in practice \cite{Kuhr2016,Ott2016,Gross2021}, and owing to less demands on resources, experimental control is often limited to small subsystems of otherwise large quantum systems, which provides an additional motivation for theoretically probing the features of quantum many-body dynamics reduced to the level of a few spins. Thirdly, a practical reason is that constructing reduced density matrices of subsystems scales exponentially with their size, but with the former two factors in mind we are content with restricting ourselves to small subsystems. 
\vspace{0.2cm}

This article is organized as follows. In Section.\ref{sec2} we give a very brief overview of the requisite notions from quantum information theory employed in this work, Section.\ref{sec3} discusses the results for the case of Ising spin chain with transverse and longitudinal fields, and Section.\ref{sec4} summarizes and concludes this article.

\section{Overview of quantum information notions} \label{sec2}

In this section, we set our conventions and briefly review the requisite notions and tools from the large subject of quantum information theory, more exhaustive coverage and discussions can be found in several books and references therein \cite{Wilde2017,Hayashi2017,Watrous2018,Holevo2019,Gour2024}. Quantum states live in the Hilbert space $\mathcal{H}$, which for quantum spin systems on $N$ sites has a tensor product structure $(\mathbb{C}^d)^{\otimes N}$ over the $N$ local Hilbert spaces each of dimension $d$ (e.g. $d\!=\!2$ for $S\!=\!1/2$ spins),  thus dim$(\mathcal{H})\!=\!d^N$.  A fundamental object of interest is the density matrix operator $\rho$ with the properties that it is Hermitian and positive semi-definite with unit trace ($\operatorname{tr}  (\rho) \!=\!1$), which means that its eigenvalues $\{ p_i\} $ can be regarded as providing a classical probability distribution. A state represented by $\rho $ is called a \textit{pure} state if $\operatorname{tr} (\rho^2) \!=\! 1$,  and \textit{mixed} if $\operatorname{tr} (\rho^2) < 1 $. Alternatively but equivalently, for a pure state one has $\rho \!=\! |\psi\rangle \langle \psi |$, whereas a mixed state is a probabilistic mixture of several pure states, $\rho \!=\! \sum_i p_i  |\psi_i\rangle \langle \psi_i |$, where the number of terms in this summation is called the rank of the density matrix. We denote by $ \mathcal{D}(\mathcal{H})$ the set of density matrices over $\mathcal{H}$. It is a convex set, meaning that any linear combination of density matrices is also a density matrix.
\vspace{0.2cm}

A quantum operation $\Lambda$ on the space of density matrix operators is called positive if it maps density matrices to density matrices, i.e., $\Lambda \!:\! \mathcal{D}(\mathcal{H}) \!\rightarrow \! \mathcal{D}(\mathcal{H})$. When however $\mathbb{I} \otimes \Lambda$ is positive, where the identity operation $\mathbb{I}$ acts on entities that are ancillary or complementary to the support of $\Lambda$, it is said to be completely-positive. Completely-positive trace-preserving (CPTP) operations/maps are popularly known as quantum channels. Partial tracing is a CPTP operation that is relevant for this work, while for e.g. partial transposition is only positive trace-preserving (PTP) but not CPTP. Such PTP operations are not physical in the sense that they can not be implemented in the laboratory, as they do not generally map density matrices to density matrices when applied to a part of the system, while CPTP maps are experimentally implementable.
\vspace{0.2cm}

Given a quantum many-body system whose Hilbert space takes the tensor product structure over its constituents, for any bipartitioning of the whole system into two subsystems $ A $ and its complement $A_c$, we have $ \mathcal{H} \!=\! \mathcal{H}_A \otimes \mathcal{H}_{A_c}$. One defines a \textit{reduced} density matrix (RDM) of the subsystem $A$ by tracing over the degrees of freedom in its complement, i.e., $\rho_A \!=\! \operatorname{tr}_{A_c} (\rho)$. For a pure state of the total system, it is said to be \textit{separable} with respect to such a bipartition if $\rho \!=\! \rho_A \otimes \rho_{A_c}$ $\big($more generally, for a mixed state of the total system, $\rho \!=\! \sum_i r_i \rho_A^{(i)} \otimes \rho_{A_c}^{(i)}$ , if the states of the subsystems are not known with certainty and form a set of possibilities indexed by $\{{i}\} $ with probabilities $r_i$$\big)$, and \textit{entangled} otherwise. The set of bipartite separable states will be denoted as $\mathcal{S}(\mathcal{H})$. There are several measures for quantifying bipartite separability and bipartite entanglement of pure states \cite{Guhne2009,HorodeckisRMP2009}, all of which are known to be equivalent to the von Neumann entropy of entanglement (of the subsystem $A$ with the rest of the system), $S_A \!=\! -\operatorname{tr} \rho_A \ln (\rho_A ) \!=\! -\sum_i p_i \ln p_i $. The second equality thus relates the von Neumann entanglement entropy (vNEE) to the classical Shannon entropy of the distribution $\{{p_i}\}$ of the eigenvalues of $\rho_A$. We shall also be interested in the second Rényi entanglement entropy (R2EE) defined as $S_{2} \!=\! -\ln \operatorname{tr}(\rho_{A})^2$, which is an experimentally measurable quantity as opposed to vNEE, due to the logarithm of the density matrix in the latter which is an experimentally ill-defined operation \cite{Islam2015}.
\vspace{0.2cm}

Let us briefly mention here that an operational meaning of vNEE is that it quantifies the number of maximally entangled spin singlet pairs that can be "distilled" from a given system in the asymptotic limit of many copies of the said system via classically communicated local operations \cite{Plenio2006,HorodeckisRMP2009,Guhne2009}. However, this theoretical limit can not be afforded by an observer in the laboratory, where one has access to only a single copy (or at most a few copies of) of the system. An operationally well-defined entropy of entanglement valid for a single copy of the system was introduced in \cite{Vidal1999} and is given by $S_{1A} \!=\! -\ln p_{max}$
where $p_{max}$ is the largest eigenvalue of $\rho_A$ (this quantity has a particular significance in quantum critical systems \cite{Eisert2005,Orus2006,Peschel2005}, and has also recently been implicated in Page curve dynamics \cite{Kehrein2024,Li2025,Jha2025}).
\vspace{0.2cm}

Mixed state bipartite entanglement refers to entanglement between two subsystems of a total system which itself is in a mixed state. This can happen when the total system is an ensemble of pure states (such as a thermal state), or when it itself is a reduction (i.e., partial trace) from a bigger pure state. We shall be concerned with the latter scenario, as our focus in this work is on quantum systems at zero temperature. Most mixed state entanglement measures require convex optimizations \cite{Plenio2006,HorodeckisRMP2009,Guhne2009} and are thus generally hard to compute in closed forms even for small systems of a few qubits. We shall therefore primarily consider a measure which is computable in closed form without requiring any optimization strategies, making it therefore the most popular measure for mixed state entanglement.

\subsection{Entanglement Negativity and Concurrence}  \label{sec2a}

An important measure of bipartite entanglement applicable to both pure and mixed states is negativity. It is based on a quantum operation called partial transposition, defined as follows. Suppose we have a quantum state $\rho_{AB}$ on a Hilbert space $\mathcal{H}_A \otimes \mathcal{H}_B$ defined on subsystems $A$ and $B$ (note that we are not assuming $AB$ to be the whole system, i.e., $\rho_{AB}$ need not be pure). Explicitly, $\rho_{AB} \!=\! \sum_{ijkl} p_{ijkl}|i\rangle\langle j| \otimes |k\rangle \langle l|$ in some chosen basis. A partial transposition with respect to $B$ is, 
\begin{equation}  \label{PT}
    \rho_{AB}^{T_B} = \underset{ijkl}{\sum} p_{ijkl}|i\rangle\langle j| \otimes |l\rangle \langle k| = \underset{ijkl}{\sum} p_{ijlk}|i\rangle\langle j| \otimes |k\rangle \langle l|   \text{    .}
\end{equation}
The positive partial transpose (PPT) criterion then asserts that if the state $\rho_{AB}$ is separable then $\rho_{AB}^{T_B}$ is also a density matrix, i.e., $\rho_{AB}^{T_B} \in \mathcal{D}(\mathcal{H}$) \cite{Peres1996,Horodeckis1996}. Consequently, if $\rho_{AB}^{T_B}$ has any negative eigenvalues then it does not represent any physical state, i.e., $\rho_{AB}^{T_B} \notin \mathcal{D}(\mathcal{H}$). We remark here that the PPT criterion is generally only a necessary condition for separability, but for systems with $d_a \times d_b \!=\! 2\times 2$ and $2 \times 3$ (a qubit-qubit or a qubit-qutrit system) it is also \textit{sufficient} \cite{Horodeckis1996}, where $d_{a (b)} \!=\! $ dim($\mathcal{H}_{A (B)}$). However, $\rho_{AB}^{T_B}$ having any negative eigenvalue is a guarantee that the system $AB$ is entangled with respect to partial transposition. 
\vspace{0.2cm}

Based on the PPT criterion, a measure of entanglement called the negativity of entanglement is defined in terms of the eigenvalues $\{p_j\} $ of $\rho_{AB}^{T_B}$ as \cite{VidalWerner2002}, 
\begin{equation}   \label{EN}
    \mathcal{N}(\rho) = \frac{1}{2} \underset{j}{\sum} \big( |p_j| - p_j \big)  \text{      .}
\end{equation}
Thus, negativity essentially enumerates the contribution of the negative eigenvalues of the partially transposed density matrix to its trace ($=1$, because partial transposition nonetheless preserves the trace). Note that owing to the closed-form expression above and the relative ease with which the partial transposition may be performed for any given system, the negativity is a readily computable measure.
\vspace{0.2cm}

Note that the partial transposition operation defined above may be written as $\Lambda_{PT} \!=\! \mathbb{I}_A \otimes \mathbb{T}_B$, where $\mathbb{T}_B$ is transposition of the chosen basis in $B$. Thus, $\Lambda_{PT}$ is in general not a CPTP map but only a PTP map, and so is not in general a physically implementable operation in the laboratory. One then might ask if measures based on this operation are experimentally measurable. We remark on this in passing that with this same concern, Ref.\cite{HorodeckiEkert2002} proposed a CPTP operation that approximates any PTP operation such as the partial transposition. This so-called structural physical approximation (SPA) of $\Lambda_{PT}$ has eventually been realized in experiments \cite{Lim2011,Bae2017}, and a counterpart to negativity for the SPA of the partial transposition map has also been defined recently \cite{Kumari2022}.
\vspace{0.2cm}

We make another remark in passing that the partial transposition $\Lambda_{PT}$ has a physical interpretation as a time-reversal operation on $B$ \cite{Busch1997,Sanpera1997} or equivalently, for a bosonic system, as a mirror reflection in the phase space \cite{Simon2000} (for fermions, the issue is more subtle \cite{Shapourian2017,Shapourian2019}). The PPT criterion in this language asserts that time-reversal on one subsystem of a separable state does not alter separability despite obviously altering the state of the said subsystem, and negativity then measures the violation of this picture whereby subsystem time-reversal yields an unphysical result no longer representable by a density matrix, thereby detecting and quantifying bipartite entanglement in a non-trivial manner.
\vspace{0.2cm}

We shall also make use of another common measure of bipartite entanglement, called as concurrence, which in particular for a mixed state of two qubits takes an easily computable closed form \cite{Wootters2001}. Given a density matrix $\rho$ describing a mixed state of two qubits, one constructs its spin-flipped version,
\begin{equation}   \label{flip}
    \bar{\rho} =  (\sigma_y \otimes \sigma_y) \rho^* (\sigma_y \otimes \sigma_y) \text{      ,}
\end{equation}
where $\rho^*$ denotes complex conjugation (in $\sigma_z$ basis). The concurrence is then defined as \cite{Wootters2001},
\begin{equation}    \label{conc}
    \mathcal{C}(\rho) = \operatorname{max}(0,\lambda_1-\lambda_2-\lambda_3-\lambda_4)  \text{      ,}
\end{equation}
where $\lambda_i (i=1,2,3,4)$ are the eigenvalues in descending order of the Hermitian matrix $\Lambda \!=\! \sqrt{\sqrt{\rho} \bar{\rho} \sqrt{\rho}} $. It is known that the negativity is upper bounded by the concurrence \cite{Verstraete2001}.

\subsection{Majorization relations}    \label{sec2b}

The notion of majorization in the mathematical theory of inequalities is one of the pillars in quantum information theory. Consider two pure states $\Psi$ and $\Phi$ of a bipartite system $A \cup B$, with corresponding pure density matrices $\rho$ and $\sigma$, and reduced density matrices of subsystem $A$ likewise denoted $\rho_A$ and $\sigma_A$ with respective vector of eigenvalues arranged in descending order being $p^{\downarrow}$ and $q^{\downarrow}$. Recall that by Schmidt decomposition, one has $\Psi \!=\! \sum_{i=1}^n \sqrt{p_i} |a_i \rangle\otimes |b_i \rangle $ and similarly for $\Phi$, in some chosen basis $\{|a_i\rangle\}$ and $\{|b_i\rangle\}$ for the subsystems $A$ and $B$. And similarly one has $\rho_A \!=\! \sum_{i=1}^n p_i |a_i\rangle\langle a_i |$ and likewise for $\sigma_A$. We say $p^{\downarrow}$ \textit{majorizes} $q^{\downarrow}$, or equivalently that $\rho_A$ majorizes $\sigma_A$, denoted as $\rho_A \succ \sigma_A$, if for each $M \in \{1,2,..,n\}$, 
\begin{equation}  \label{maj}
    \sum_{i=1}^M p^{\downarrow}_i \geq \sum_{i=1}^M q^{\downarrow}_i  \text{  ,}
\end{equation}
with equality necessarily at $M=n$. 
\vspace{0.2cm}

Now, an important theorem due to \cite{Nielsen1999,NielsenVidal2001} links majorization with bipartite entanglement by asserting that $\rho_A$ majorizes $\sigma_A$ if and only if the state $\Phi$ contains as much or more entanglement amongst its subsystems $A, B$ than the state $\Psi$. Equivalently, $\sigma_A$ is \textit{more} mixed than $\rho_A$ \cite{NielsenVidal2001}. This imposes a one-way partial-order among quantum states and is thus an entanglement monotone \cite{Vidal2000}. This partial order is based on generic protocols for conversion of one state to another by with local operations (such as measurements, unitaries and any other completely-positive operations) that are classically communicated between observers carrying out the said operations (see for e.g. \cite{Hayashi2017,Gour2024}. Because the operations are local, they can not increase entanglement between the states held by the observers, and thus any legitimate measure of entanglement needs to be a non-increasing monotone under local operations and classical communications (LOCC). Majorization relations are one such family of monotones, and thus the state $\Psi$ can be converted to $\Phi$ but not vice-versa. Besides, unlike pure state bipartite entanglement quantifiers such as the von Neumann entropy which quantify LOCC convertibility between asymptotically large number of copies of the two states and thereby quantify distillation of maximally-entangled singlets from a higher entangled state resulting in a lower entangled state in this asymptotic limit, giving a single inequality governing this situation, majorization relations are a stronger and more detailed set of criteria in the sense of providing providing several inequalities governing this problem \cite{NielsenVidal2001}.

\subsection{Quantum mutual information}  \label{sec2c}

Given a quantum system $AB$ composed of subsystems $A$ and $B$, the quantum bipartite mutual information (BMI) between them is defined in terms of von Neumann entropies as,
\begin{equation}   \label{BMI}
    \mathcal{I}_2(A \mathord{:} B) = S_A + S_B - S_{AB}   \text{  .}
\end{equation}
It is always non-negative, quantifies classical as well as quantum correlations between $A$ and $B$, or in other words it quantifies in a sense the total information that $A$ has about $B$ and vice-versa \cite{Groisman2005}, and upper-bounds connected two-point correlation functions of local observables in the ground states of quantum many-body systems \cite{Wolf2008}. It vanishes if and only if the state $\rho_{AB}$ factorizes as $\rho_A \otimes \rho_B$.
\vspace{0.2cm}

One can also formally define mutual information between more than two systems. Of particular interest is the quantum tripartite mutual information (TMI) defined for a system $ABC$, 
\begin{equation} \label{TMI}
    \mathcal{I}_3(A \mathord{:} B \mathord{:} C) = \mathcal{I}_2(A \mathord{:} B)+\mathcal{I}_2(A \mathord{:} C)-\mathcal{I}_2(A \mathord{:} BC).
\end{equation}
It measures the difference of (mutual) information about the subsystem $A$ that subsystems $B$ and $C$ have individually and what they together ($BC$) know about $A$. Speaking in terms of correlations, the TMI quantifies the correlations that $BC$ has with $A$ which is not present in the cumulative correlations with $A$ that $B$ and $C$ have individually. An intriguing fact is that the TMI can be negative, which is interpreted in such cases as the subsystem $BC$ having more information about $A$ than $B$ and $C$ have individually. In the context of non-equilibrium dynamics, this is often interpreted as (a signature of) \textit{scrambling} of localized information (about $A$, now imagined as a reference system) in $B$ and $C$ to a more non-local sharing of that information into $BC$. Thus the negativity of TMI has been proposed as a measure of quantum information scrambling \cite{Hosur2016} and it has received some attention in non-equilibrium quantum dynamics literature \cite{Iyoda2018,Seshadri2018,Schnaack2019,Kuno2022,Sur2022,Monaco2023,Monaco2023a,Caceffo2023,Maric2023,Maric2023a,Maric2023b}. It is worth mentioning that TMI is also directly related to topological entanglement entropy \cite{Kitaev2006,Levin2006} of topologically-ordered phases of quantum matter and has also been proposed as a measure of tripartite entanglement in a four-partite quantum system \cite{Hosur2016}.

\section{Results for the mixed-field Ising spin chain}              \label{sec3}

We demonstrate a bulk of our results for the paradigmatic Ising spin chain in the presence of both transverse and longitudinal fields, given by the Hamiltonian, 

\begin{equation}     \label{TFIM}
    \mathcal{H}_I = -J\sum_{j=1}^{N-1}\sigma^z_j \sigma_{j+1}^z - h_x\sum_{j=1}^N \sigma_j^x - h_z\sum_{j=1}^N \sigma_j^z  \text{   
        .}
\end{equation}
The model is integrable when $h_z\!=\!0$, with the Ising critical point at $J\!=\!h_x$ separating the symmetry-broken ferromagnetic/antiferromagnetic phases from the paramagnetic phase \cite{Pfeuty1970}. It is non-integrable when $h_z \! \neq \! 0$, and the kink-antikink excitations in the symmetry-broken phases (ferromagnetic or anti-ferromagnetic) are confined \cite{McCoy1978,Delfino1996,Fonseca2003,Rutkevich2008} and consequently exhibits anomalously slow thermalizing dynamics and other associated features after a quench to this non-integrable regime \cite{Banuls2011,Kormos2016,Lin2017,Alvaredo2020,Liu2019,Robinson2019,Scopa2022,Birnkammer2022,Knaute2023,Kaneko2023,Robertson2024}. In Ref.\cite{Banerjee2025}, we presented numerical evidence of information backflow and non-Markovianity of the dynamics of small subsystems when quenched from a paramagnetic ground state to the ferromagnetic side, whereas the reversed quench showed practically Markovian dynamics. This section deals with dynamical behaviour of the quantum informational quantities introduced earlier, which for the case of paramagnetic-to-ferromagnetic quench will be seen to have distinctive dynamical features (that can be considered to accompany subsystem non-Markovianity), whereas rather featureless dynamical behaviour will be seen in the reversed quench.
\vspace{0.2cm}

We used the matrix product states framework for our simulations, implemented with the ITensors library in Julia \cite{Itensor}. Results in this article were obtained with time-steps $\tau\!=\!0.01$ seconds for our TEBD2 simulations (thus, numerical errors were $\mathcal{O}(10^{-4})$), and we verified that the results were unchanged with time-steps of $\tau\!=\!0.002$ seconds. The total number of spins was fixed at $N\!=\!200$, and results were verified to be independent of the other system sizes. The cutoffs for MPS truncations were fixed at $10^{-9}$, and maximum bond-dimensions were fixed at $50$. The ground state wavefunctions in MPS form were obtained with the density matrix renormalization group (DMRG) algorithm \cite{Schollwoeck2011}, also implemented with the same parameters as mentioned above. 

\begin{figure}[tp]
    \centering
    \includegraphics[width=5.1cm,height=4cm]{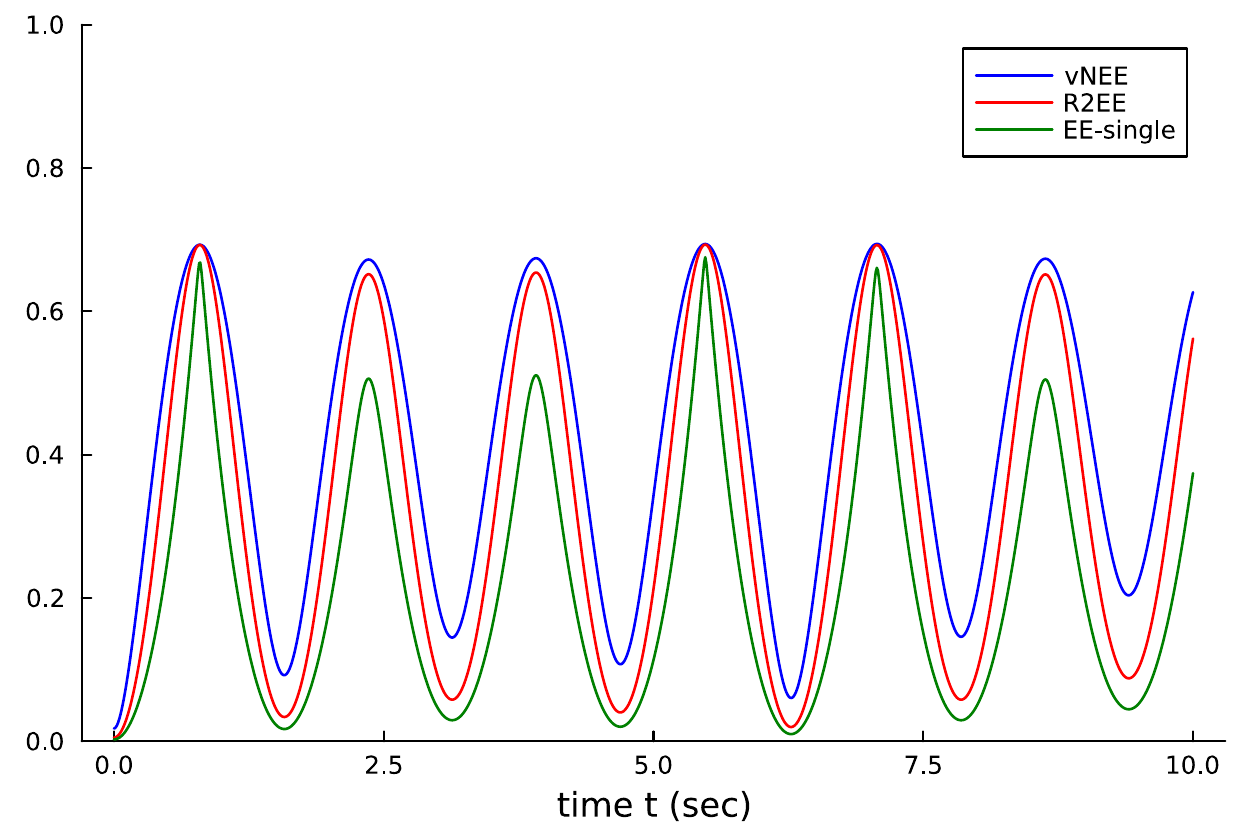}
    %\hspace{0.5cm}
    \includegraphics[width=5.1cm,height=4cm]{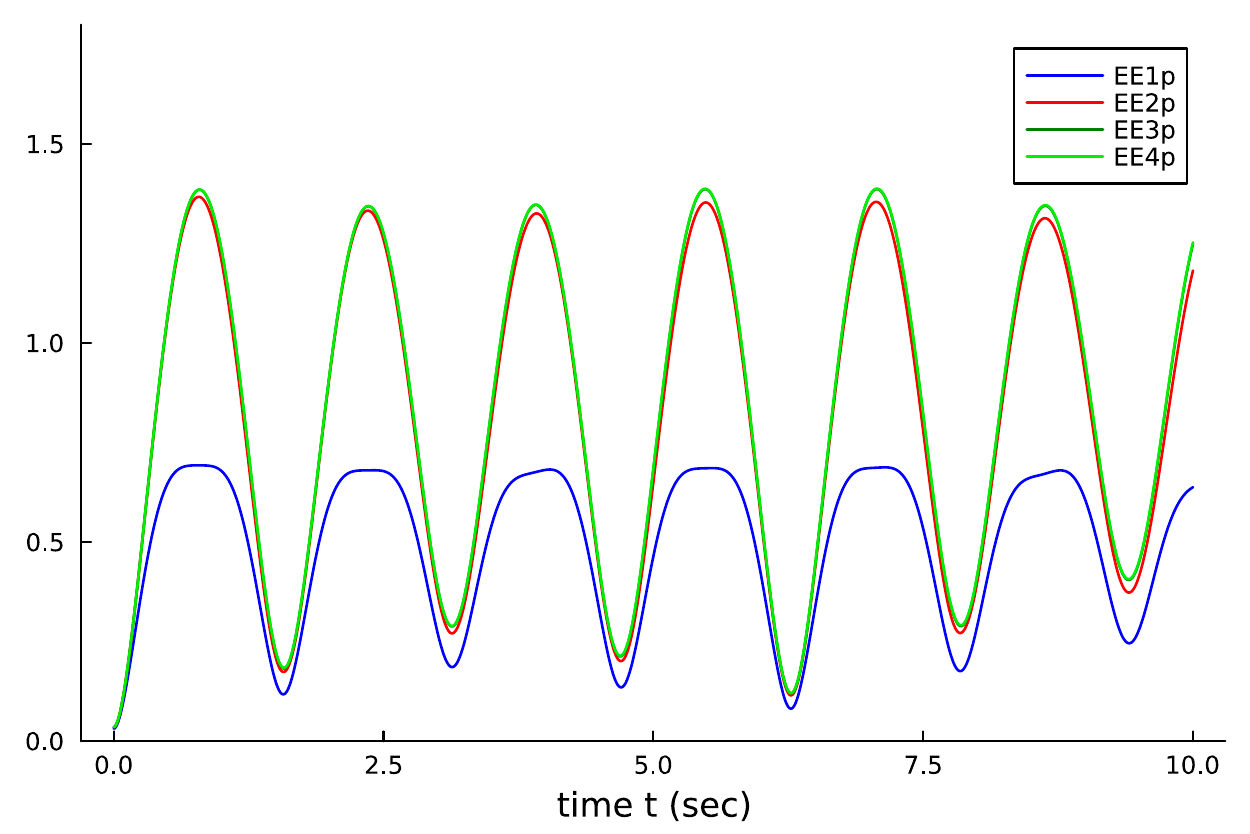}
    %\hspace{0.5cm}
    \includegraphics[width=5.1cm,height=4cm]{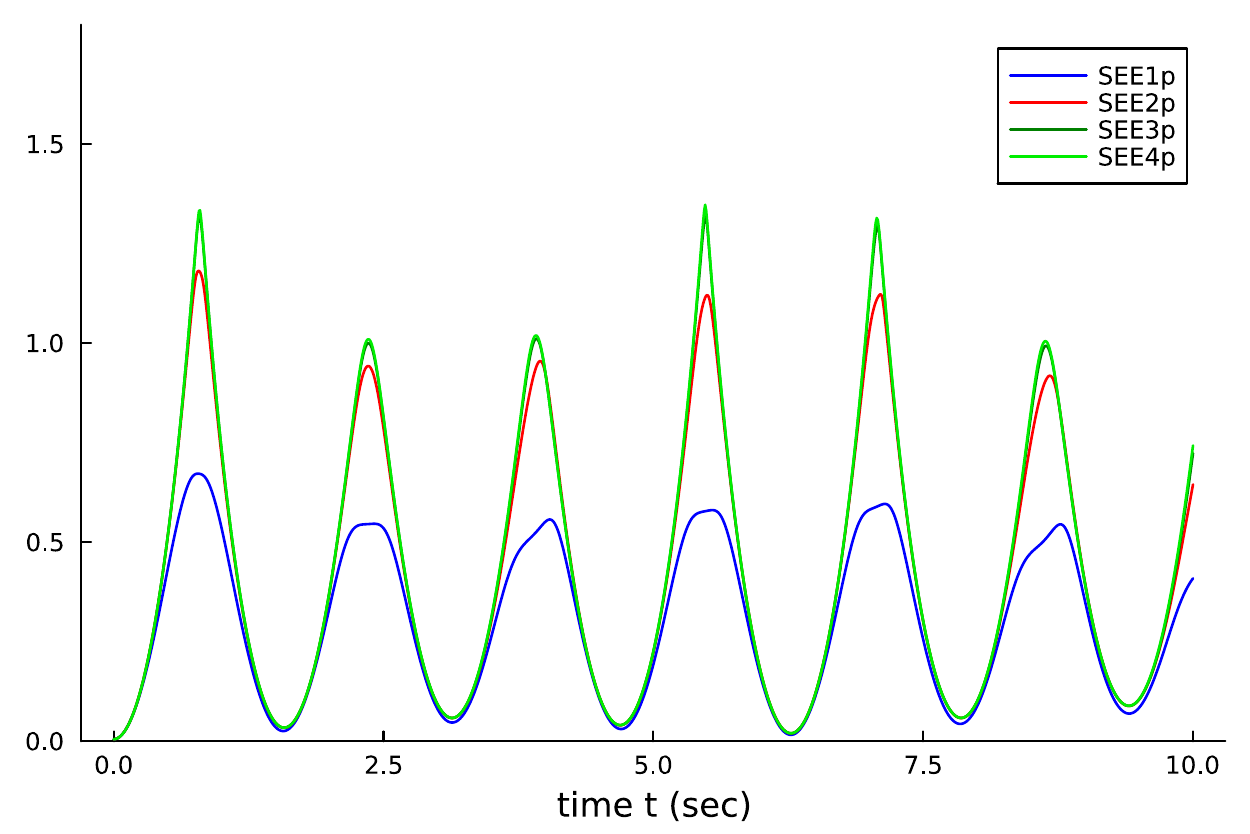}   
    
    \caption{\fontsize{10}{12} \selectfont Dynamical behaviour of bipartite pure state entanglement entropies (y-axes) for the quench protocol $(J,h_x,h_z) \!=\! (0.2,1,0) \rightarrow  (1,0.1,0.5)$. (\textbf{Left}) Half-chain von Neumann, Renyi-2 and single-copy entanglement entropies, denoted respectively as vNEE, R2EE, EE-single in the figure ; (\textbf{center}) von Neumann entanglement entropies of small subsystems of up to four-spins, where for e.g. EE1p refers to the one-spin subsystem and likewise for the others; (\textbf{right}) Single-copy entanglement entropies of small subsystems of up to four spins, where for e.g. SEE1p refers to the one-spin subsystem and likewise for the others. Highly oscillatory behaviour is seen in all these quantities, with a time-scale of $\sim 1.55$ seconds between consecutive minima (or maxima), and $\sim 0.78$ seconds between a pair of consecutive minimum and maximum. Note the appearance of non-analytic cusps in single-copy entanglement entropies.}
	\label{fig:fig1}
\end{figure}

\subsection{Paramagnetic to ferromagnetic quench} We first show results for the case of quenching from the paramagnetic ground state at $(J,h_x,h_z) \!=\! (0.2,1,0)$ to the ferromagnetic side $(J,h_x,h_z) \!=\! (1,0.1,0.5)$. Simulation times $t$ are in units of $J^{-1}$, denoted in the figures by (sec) in the x-axes.

\subsubsection{\textbf{Entanglement entropies}}

In Fig.\ref{fig:fig1} we show the dynamical behaviour of pure state bipartite entanglement entropies for half-chain as well as small subsystems of up to four spins, as measured by von Neumann and Renyi-2 entanglement entropy as well as single-copy entanglement entropies. The highly (and seemingly persistent) oscillatory behaviour of (von Neumann and Renyi) entanglement entropies in this quench, attributed to the slow growth of entanglement due to confinement in the non-integrable ferromagnetic side, is known from the literature and is being shown here for completeness. It is worth noting that the maxima and minima of entanglement entropies of the various considered subsystems occur at almost the same instants of time, approximately irrespective of the size of the subsystem, and there exists a time-scale of $\sim 1.55$ seconds between consecutive minima (or maxima), and $\sim 0.78$ seconds between a pair of consecutive minimum and maximum. 
\vspace{0.2cm}

We also take notice of the non-analytic cusps in the dynamics of single-copy entanglement entropies of half-chains and subsystems of three and four spins (but not of one and two spins). These cusps are due to avoided crossings (or approximately so) between the largest two eigenvalues of the corresponding reduced density matrices, and the first avoided crossing can overlap with the first point (in time) of a dynamical quantum phase transition in this same paramagnetic-to-ferromagnetic quench \cite{Nicola2021}. See also \cite{Li2025,Kehrein2024} in which cusps in the same quantity (named there as the min-entropy) appear in the Page curve of Hamiltonian dynamics, but only in the thermodynamic limit unlike the present case. In that case, the time at which a cusp in this quantity appears indicates the point of maximal pure state bipartite entanglement after which the system starts "cooling" off and is considered to signify a first-order-like phase transition in the Page curve \cite{Page1993}. Moreover, given that the largest eigenvalue of a density matrix corresponds to the ground state of the corresponding entanglement Hamiltonian \cite{Dalmonte2022}, which in a sense is an effectively reduced Hamiltonian corresponding to the mixed state denoted by the said density matrix, non-analyticities in the logarithm of the largest eigenvalue corresponds to non-analytic changes (such as phase transitions) in the ground state space of the corresponding entanglement Hamiltonian \cite{Kehrein2024}.

\begin{figure}[tp]
    \centering
    \includegraphics[width=5.1cm,height=4cm]{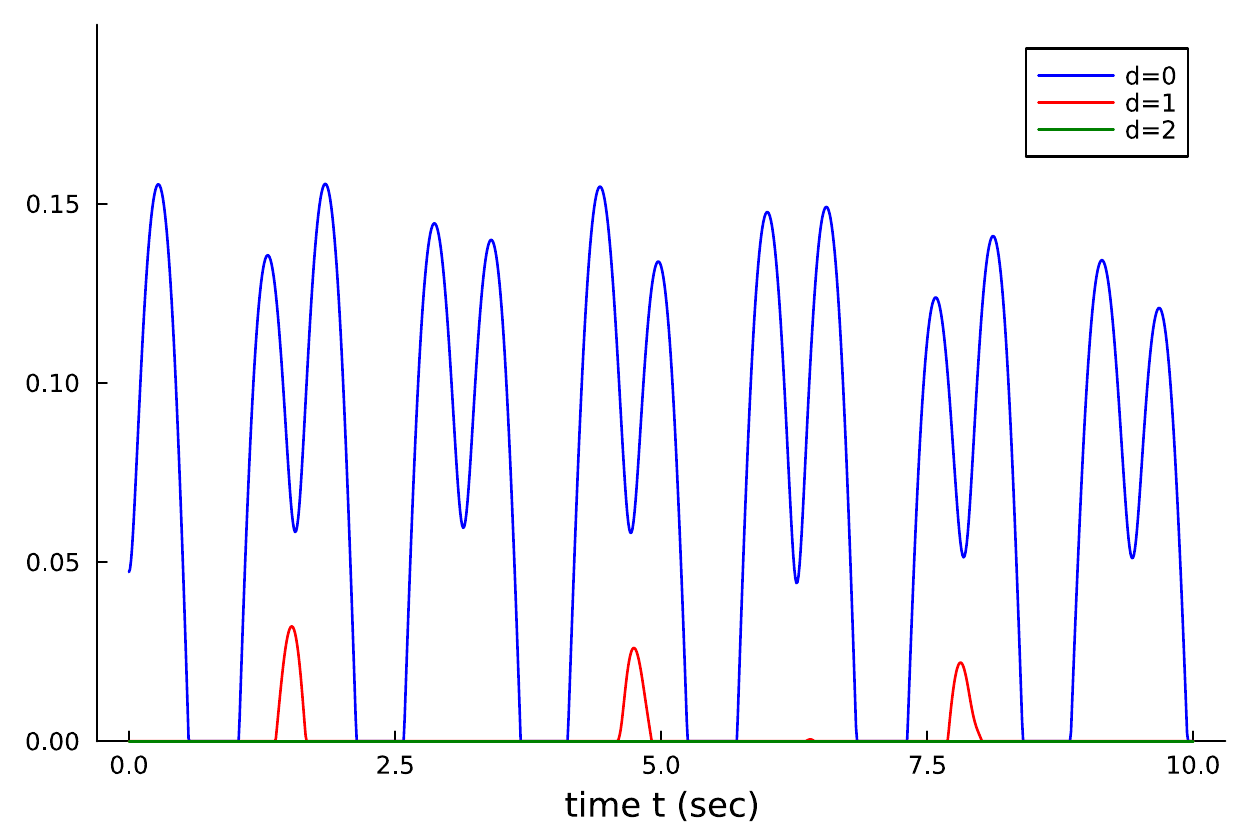}
    %\hspace{0.5cm}
    \includegraphics[width=5.1cm,height=4cm]{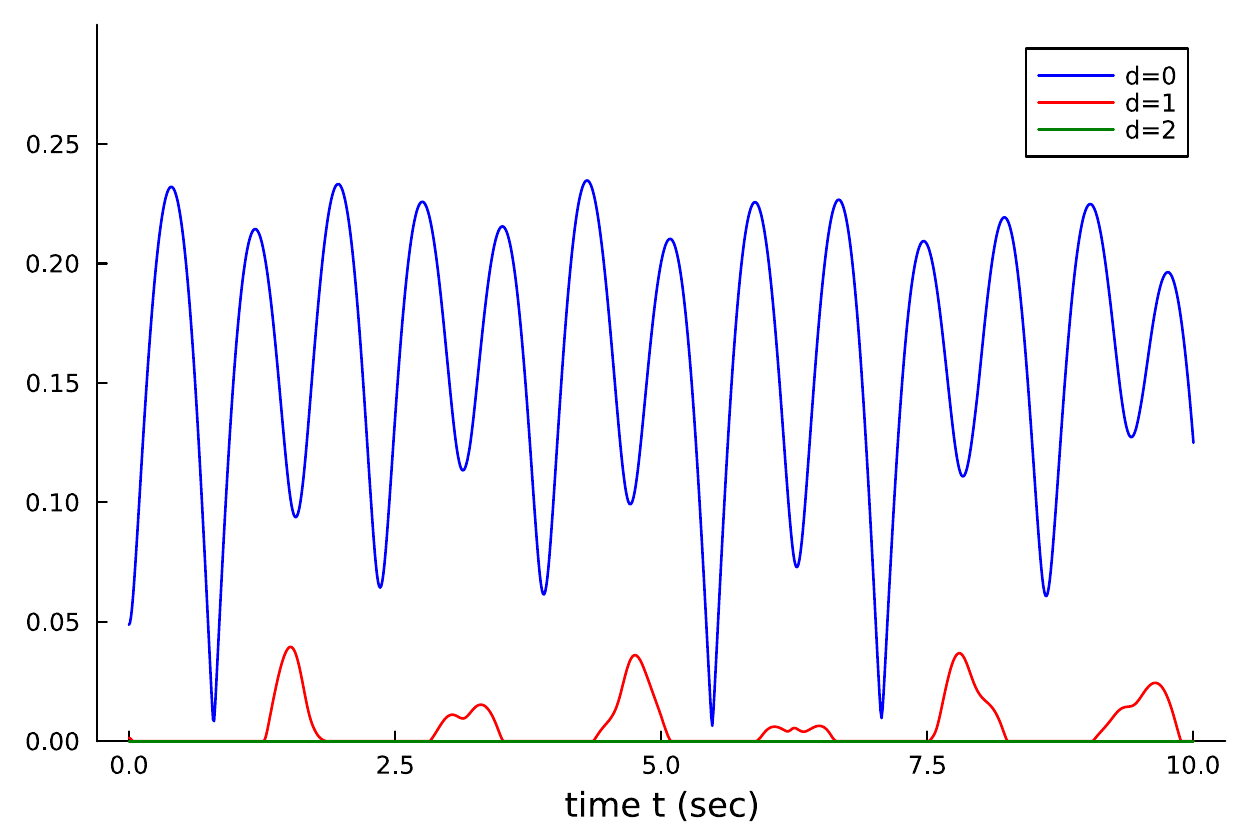}
    %\hspace{0.5cm}
    \includegraphics[width=5.1cm,height=4cm]{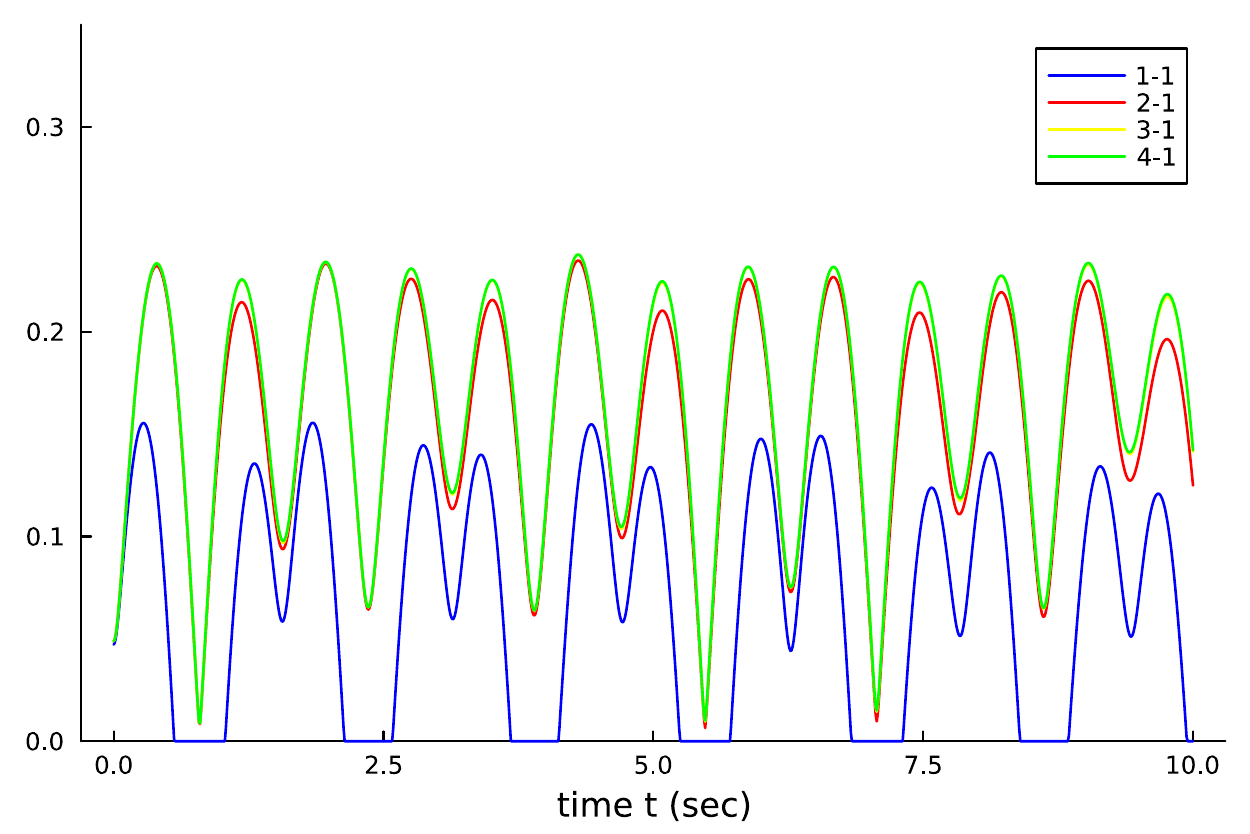}

    \caption{\fontsize{10}{12} \selectfont Dynamics of entanglement negativity (y-axes) for the quench protocol $(J,h_x,h_z) \!=\! (0.2,1,0) \rightarrow  (1,0.1,0.5)$. (\textbf{Left}) For spin-spin pair, (\textbf{center}) two-spin with a third spin at separation $d$ (\textbf{center}), and (\textbf{right}) contiguous blocks of $(n-1)-$spins with the $n^{th}-$spin. In the last figure, for e.g. $2-1$ refers to the case $n\!=\!3$.}
	\label{fig:fig2}
\end{figure}

\subsubsection{\textbf{Entanglement negativity of small subsystems}}

In Fig.\ref{fig:fig2} we show the behaviour of entanglement negativity (Eq.\ref{EN}) between two spins (Fig.\ref{fig:fig2}, left) at separations denoted by $d$, between two contiguous spins and a third spin (Fig.\ref{fig:fig2}, center) at separations denoted by $d$, and between the first $(n-1)$ spins and the last spin of a contiguous block of $n$ spins (Fig.\ref{fig:fig2}, right) with $n\!=\!\{2,3,4,5\}$. Remarkably, a sudden "death of entanglement" effect [refs], occurs already for two-spins adjacent ($d\!=\!0$) and next-nearest-neighbor ($d\!=\!1$) to each other at regular time intervals in this quench dynamics, although the entanglement in the latter case is already quite small. This is also seen for the case of negativity between two contiguous spins with a third spin at separation $d\!=\!1$, albeit again the entanglement here too is small. However, no such phenomena is seen for the entanglement between contiguous blocks of $(n-1)$ spins with the $n^{th}-$spin for $n\!=\!\{3,4,5\}$. This series of sudden deaths and revivals of entanglement between two adjacent spins is an interesting phenomenon of paramagnetic-to-ferromagnetic quench dynamics, and to the best of our knowledge, has not been uncovered in the literature of non-equilibrium dynamics of closed quantum many-body systems (see however \cite{Lopez2008,Rau2008,Yu2009}). To ensure that this is not an artifact of employing negativity as the measure of mixed state entanglement, we show the behaviour of concurrence for the mixed state entanglement of two adjacent spins in Fig.\ref{fig:fig3}, where also this series of sudden deaths and revivals of entanglement is observed.
\vspace{0.2cm}

Recall from our discussion in section \ref{sec2a} that zero negativity (i.e., satisfying the PPT criteria) is sufficient for separability of a two-spin state, but this is not the case for other higher dimensional subsystems such as the three-spin subsystem with a separation $d\!=\!1$ between the first two spins and the third spins, which shows zero negativity at certain time intervals (center, Fig.\ref{fig:fig2}), but despite of this there may still be entanglement between them. This form of entanglement with zero negativity is called bound entanglement or PPT-entanglement, thus named because maximally entangled singlets can not be distilled from such states via LOCC in a quantum informational experiment/protocol, and understanding them continues to be a perplexing problem (moreover, bound-entangled states can also have non-zero negativity) \cite{Hiesmayr2024}. Thus, we shall refrain at this time from claiming that the zeros of negativity in the latter case above signify separability at these time intervals, but only that they indicate undistillability via LOCC protocols.

\subsubsection{\textbf{Recurrent Page-like dynamics of one- and two-spin subsystems}} 

\begin{figure}[tp]
    \centering
    \includegraphics[width=8cm,height=6cm]{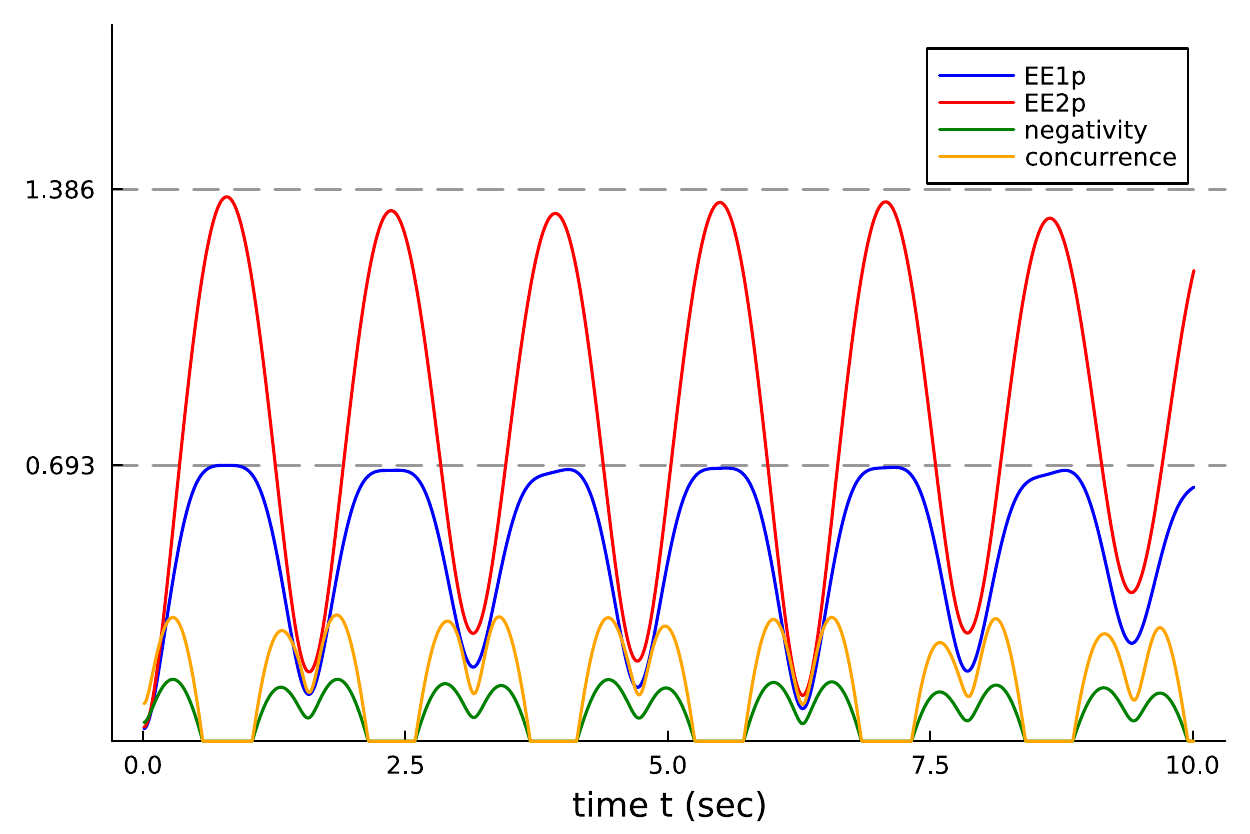}
    \caption{\fontsize{10}{12} \selectfont Comparison of pure state and mixed state bipartite entanglement for a two-spin subsystem, for the quench protocol $(J,h_x,h_z) \!=\! (0.2,1,0) \rightarrow  (1,0.1,0.5)$. The grey dashed lines are $\operatorname{ln}(2) \sim 0.693$ and $\operatorname{ln}(4) \sim 1.386$.}
	\label{fig:fig3}
\end{figure}

A rather remarkable observation can be made in Fig.\ref{fig:fig3}, which shows a comparison of the pure state and mixed state bipartite entanglement for a two-spin subsystem, as measured by the von Neumann entropy in the former, and negativity and concurrence in the latter. It is seen that it is precisely within the time intervals in which mixed state entanglement is zero (i.e., the two-spin pair is separable) that the pure state entanglement attains its maxima. Moreover, the maxima of one-spin and two-spin entanglement entropy (denoted by EE1p and EE2p in Fig.\ref{fig:fig3}, respectively) is attained at values close to their theoretical maximum of $\operatorname{ln}(2) \!\sim\! 0.693$ and $\operatorname{ln}(4) \!\sim\! 1.386$ respectively (recall that the von Neumann entropy of a density matrix of dimension $D$ is upper bounded by $\operatorname{ln}(D)$; for a system of $n$ spin-$1/2$ spins, $D\!=\! 2^n$). This theoretical maximum is attained when the density matrix is maximally mixed (and therefore diagonal), i.e., all its eigenvalues $p_i \!=\!\mathbb{I}/D, \hspace{0.1cm} \forall i$.  Naturally, a two-spin (reduced) density matrix that is diagonal has trivial partial transpose, and thus its negativity is zero and consequently the pair is unentangled. This is what is being seen with the overlap of vanishing negativity (and concurrence) with the almost maximal EE2p in Fig.\ref{fig:fig3}. The maximum (or nearly so) bipartite pure state entanglement of a spin and a spin-spin pair with the rest of the system happens when the former is unentangled with its neighboring spin and the constituent spins in the latter pair are unentangled with each other. Furthermore, EE1p and EE2p attain (values close to) their theoretical maximum (of $\operatorname{ln}(2)$ and $\operatorname{ln}(4)$ respectively) repeatedly, but that is not the case with von Neumann entropies of three or more spin subsystems, whose maxima are well below their respective theoretical maximum, as seen in Fig.\ref{fig:fig1} (center). This attainment of nearly maximal value of EE1p and EE2p and the subsequent declines to much lower valued minima is reminiscent of Page curve dynamics of bipartite quantum entanglement \cite{Page1993}, with the extra feature here that this happens repeatedly in an almost-periodic manner and only for one- and two-spin subsystems (it is also intriguing to make a connection here with a result in \cite{Banerjee2025} that one- and two-spin subsystems showed the strongest signatures of backflows of information and non-Markovianity). Page curve dynamics of entanglement has recently received much interest in non-equilibrium dynamics of quantum many-body systems, both closed \cite{Kehrein2024,Li2025,Jha2025} and open \cite{Su2021,Dadras2021,Chen2021,Glatthard2024,Glatthard2025,Ganguly2025,Ray2025} (Ref.\cite{Su2021} also suggested a connection between Page curve and non-Markovianity in a different context), and it would be very interesting to study the emergence of the recurrent Page-like dynamics seen in this work in other contexts.
\vspace{0.2cm}

Moreover, the one-spin subsystem stays at or near its maximally mixed state for a slightly extended period of time (hence the plateaus in EE1p) each time it reaches this maximal state. Thus, in this dynamics, a periodic albeit short-lived attainment of approximately $1-$uniform state occurs (a $k-$uniform state of, say, spins is a state in which all $k$-spin reduced density matrices are maximally mixed, and for $k>1$ has powerful implications for autonomous quantum error correction \cite{Scott2004}), which is nothing but the single-spin Greenberger–Horne–Zeilinger (GHZ) state (but this is not true for $k>1$). All of these features are retained and somewhat strengthened deeper into the non-integrable regime (higher values of $h_z$). A microscopic analytical description of these features valid in the deeply non-integrable regimes would be very useful to investigate in future.

\begin{figure}[tp]
    \centering
    \includegraphics[width=6cm,height=5cm]{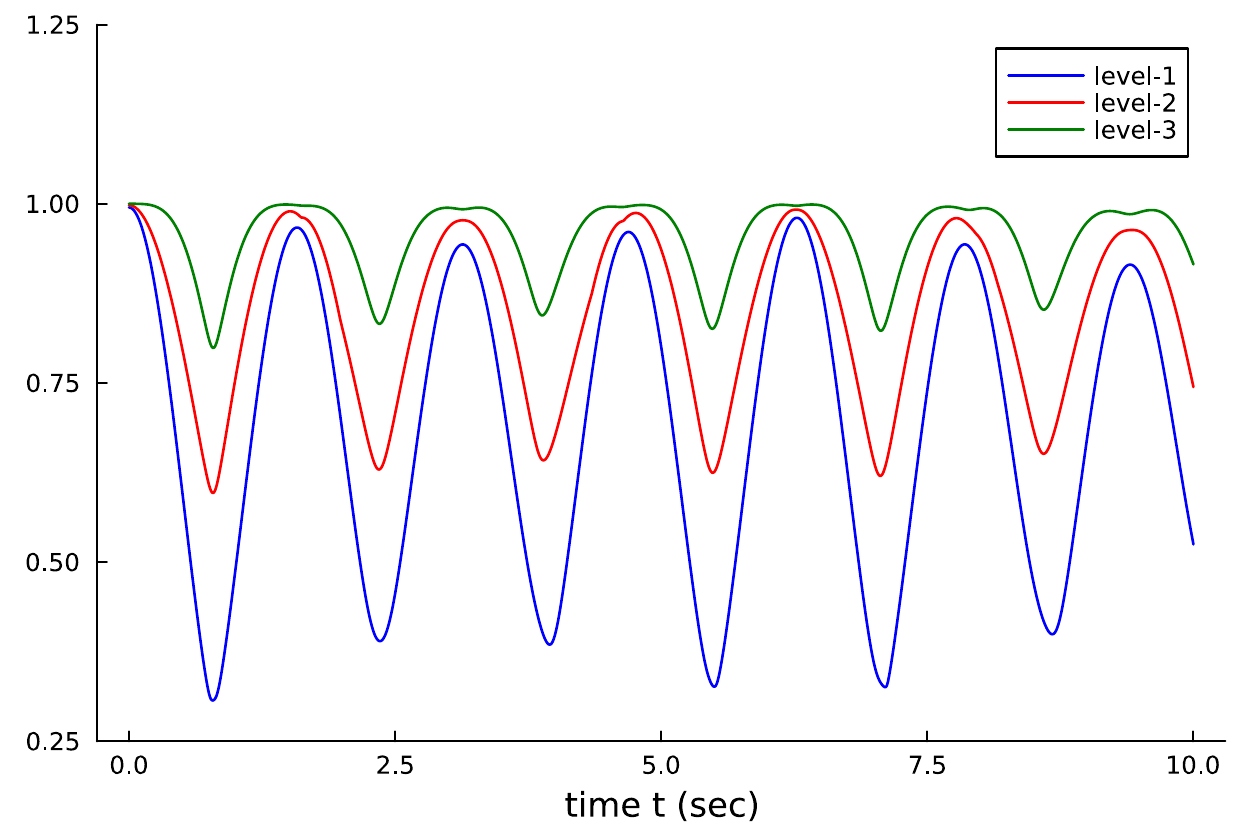}
    \hspace{0.5cm}
    \includegraphics[width=6cm,height=5cm]{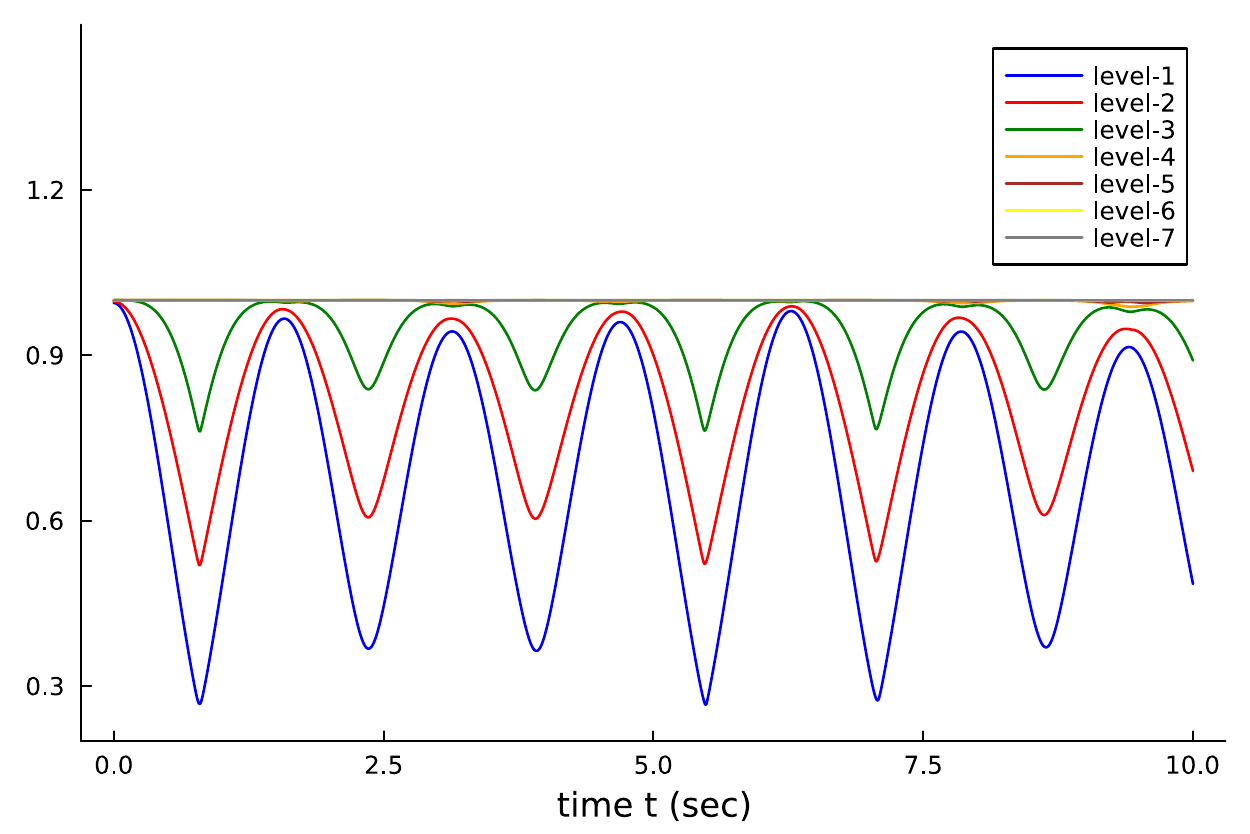}
    \caption{\fontsize{10}{12} \selectfont Dynamics of majorization relations for two-spin (\textbf{left}) and three-spin (\textbf{right}) subsystems, for the quench protocol $(J,h_x,h_z) \!=\! (0.2,1,0) \rightarrow  (1,0.1,0.5)$. The level-$k$, plotted in the y-axes, refer to the sums of first $k$ eigenvalues of subsystems' reduced density matrices, arranged in descending order. These show a non-monotonic and highly oscillatory behaviour, and a time-scale of $\sim 1.55$ seconds between consecutive minima or consecutive maxima, and $\sim 0.78$ seconds between a pair of consecutive minimum and maximum.}
	\label{fig:fig4}
\end{figure}

\subsubsection{\textbf{Majorization relations}} Next we turn to the dynamics of majorization relations discussed in section \ref{sec2b}, shown in Fig.\ref{fig:fig4} for two-spin and three-spin subsystems (larger subsystems show similar behaviour). In the figure, by "level-k" we mean the sum of the first $k$ largest eigenvalues of the reduced density matrix corresponding to the subsystem in question. The oscillatory behaviour seen in Fig.\ref{fig:fig4} for both subsystems is perhaps unsurprising at this point, given the same behaviour seen in the entanglement entropies (in particular the single-copy entanglement entropies (Fig.\ref{fig:fig1}, right) which directly manifests the oscillatory behaviour of the largest eigenvalue, shown in blue in Fig.\ref{fig:fig4}). 
\vspace{0.2cm}

This highly non-monotonic dynamics of the eigenvalue-sums shows that the majorization inequalities (Eq.\ref{maj}) are not satisfied at all times but only between pairs of consecutive maxima and a minima. Thus, the "mixedness" of the subsystems' reduced density matrices is non-monotonic, and this is a signature of the slowly relaxing and non-thermalizing (or inefficiently thermalizing) nature of this non-integrable quench dynamics, because a fast and efficient thermalization dynamics results in any and all subsystems becoming increasingly mixed sufficiently quickly so that ultimately the system as a whole may be described by a Gibbs thermal ensemble (note that this does not mean that all subsystems would eventually be maximally-mixed at thermalization, which is a special case possible only at infinite temperature). This is also evident in Fig.\ref{fig:fig4} (bottom) for three-spin subsystems, where only the first three largest eigenvalues matter and already from the fourth onward the level-k sums ($k \geq 4$) equal practically $1$ (i.e., fifth, sixth and seventh eigenvalues are negligible all across the simulation times), which is a signature of insufficient mixedness of this subsystem. Moreover, again a time-scale of $\sim 1.55$ seconds between consecutive minima or consecutive maxima, and $\sim 0.78$ seconds between a pair of consecutive minimum and maximum can be identified.
\vspace{0.2cm}

Thinking also in terms of this quench dynamics as a platform for quantum informational experiments, the results in Fig.\ref{fig:fig4} mean that only between two consecutive extrema are the instantaneous states of the whole system convertible via LOCC operations (see the discussion in section \ref{sec2b}, below Eq.\ref{maj} and also \cite{Amico2022}). However, the direction of this convertibility reverses after each extrema. For example, between the initial time and the first minimum in either of the subsystems in Fig.\ref{fig:fig4}, the level-$(1,2,3)$ sums are decreasing with time, implying LOCC convertibility from $\Psi(t) \!\rightarrow\! \Psi(t-\tau)$ at each time-step, where $\Psi(t)$ is the full system state at time $t$. However, this direction of convertibility reverses between the first minimum and the first maximum, which allows LOCC convertibility from $\Psi(t) \!\rightarrow\! \Psi(t+\tau)$. This is consistent with first an increase and then decrease of pure state bipartite entanglement in Fig.\ref{fig:fig1}.
\vspace{0.2cm}

Moreover, it may be expected that under the action of any Markovian process, a quantum system should show a monotonic non-increasing behaviour of level-$k$ sums, and therefore majorization relations should be respected at all steps of the said Markovian process (we are not aware of a general proof of this valid for any Markov process, see however \cite{Latorre2005,Orus2005} (for majorization along RG flows, which are Markov processes by definition), \cite{Yuan2010} (for Markovian open quantum systems) and \cite{Lostaglio2022} (for Markovian thermal processes) for related discussions in different contexts). We conjecture that the highly non-monotonic behaviour of these quantities are also signatures of the subsystems' dynamics being non-Markovian, adding to the discussion presented in \cite{Banerjee2025}. It would be worthwhile in future to investigate in detail the direction of implication between Markovianity and majorization in general.

\begin{figure}[tp]
    \centering
    \includegraphics[width=5.1cm,height=4cm]{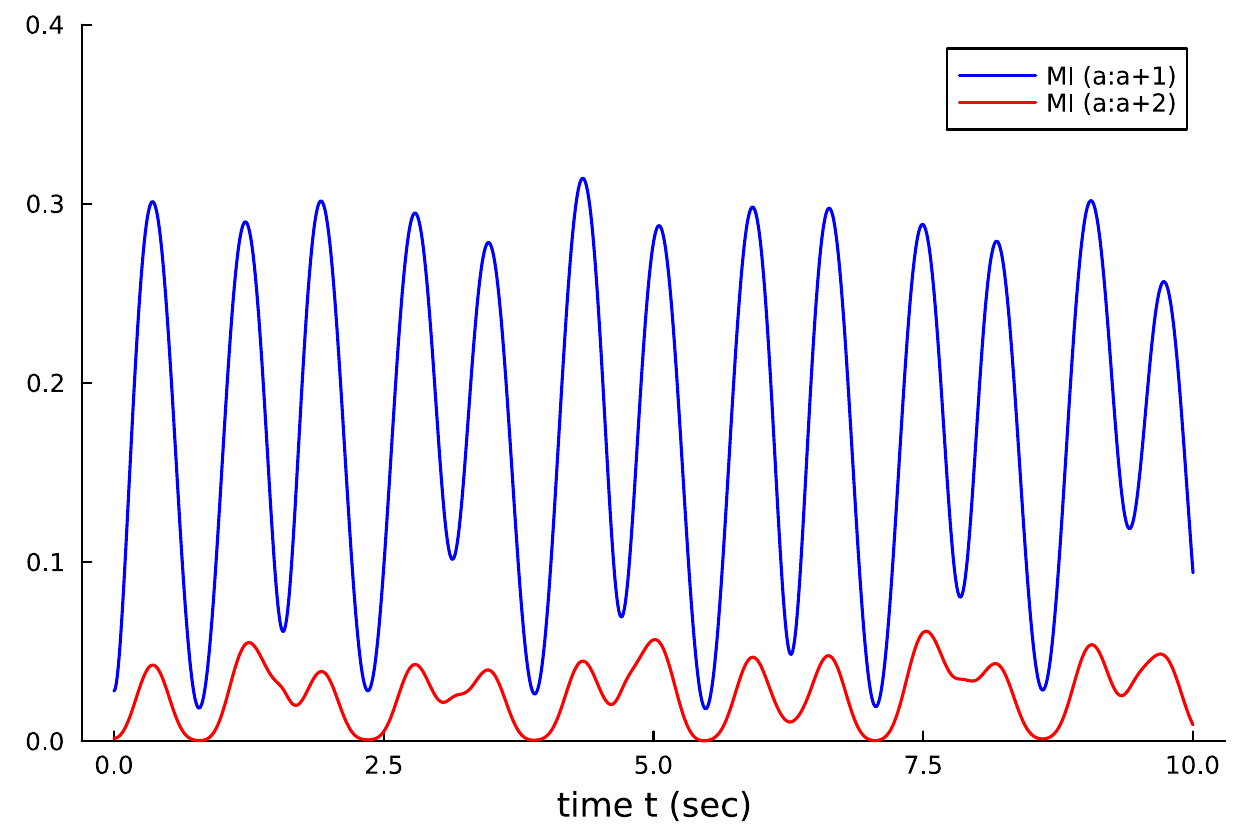}
    %\hspace{0.5cm}
    \includegraphics[width=5.1cm,height=4cm]{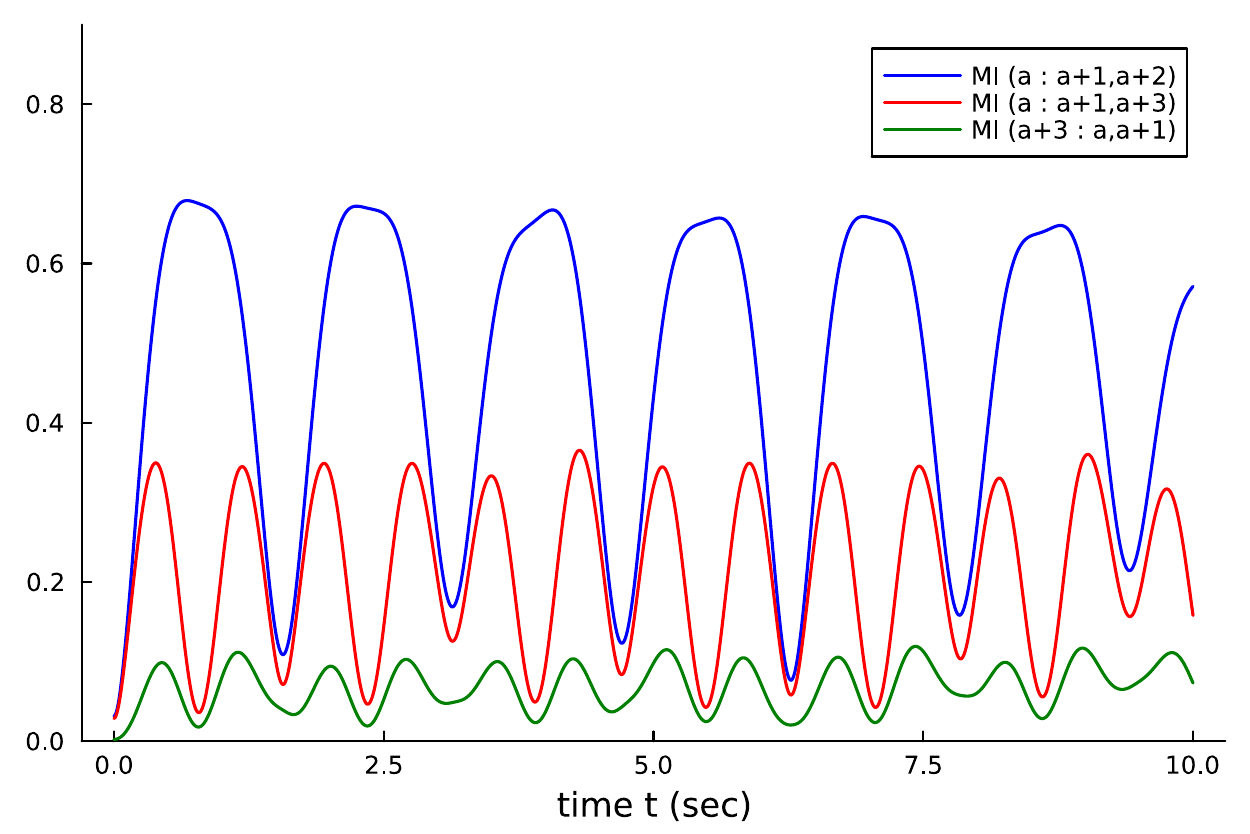}
    %\hspace{0.5cm}
    \includegraphics[width=5.1cm,height=4cm]{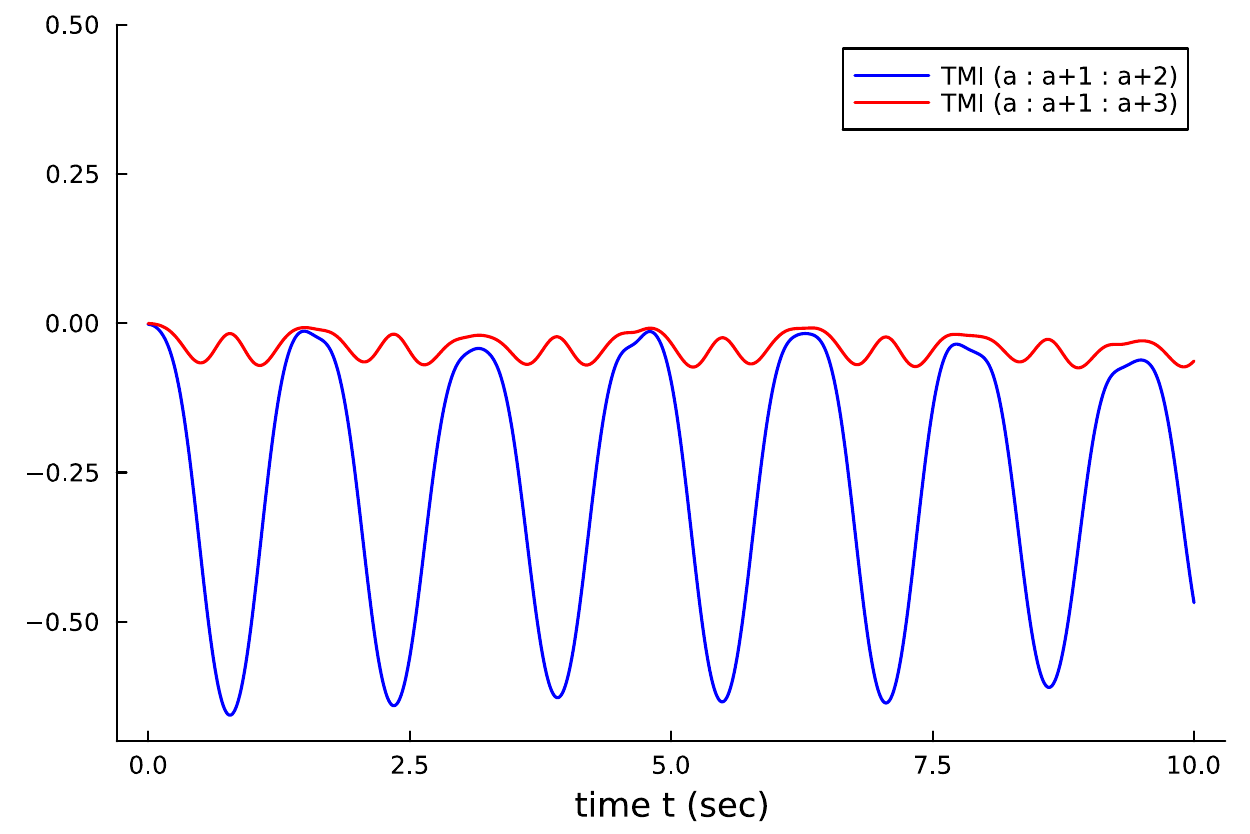}
    \caption{\fontsize{10}{12} \selectfont Dynamical behaviour of bipartite mutual information between two spins (\textbf{left}), between a spin and another pair of spins (\textbf{middle}), and tripartite mutual information between three spins (\textbf{right}), in the quench protocol $(J,h_x,h_z) \!=\! (0.2,1,0) \rightarrow  (1,0.1,0.5)$. $a$ denotes the center of the spin chain, but the results are not dependent on this choice as long as we are away from the boundaries of the spin chain due to translational invariance. Note in the middle figure that the maxima of MI($a:a+1,a+2$) and odd-numbered minima (first, third,...) of MI($a:a+1,a+3$) occur at approximately the same points of time, whereas the minima of the former and even-numbered minima of the latter approximately occur together. A timescale of $\sim 1.55-1.56$ seconds appears again between consecutive minima or consecutive maxima in MI($a:a+1,a+2$) and TMI($a:a+1,a+2$), and half of that ($\sim 0.77-0.78$ seconds) between the consecutive minima or consecutive maxima of MI($a:a+1,a+3$).}
	\label{fig:fig5}
\end{figure}

\subsubsection{\textbf{Local mutual information dynamics and scrambling-unscrambling}} Next we turn to the behaviour of bipartite mutual information and tripartite mutual information (TMI) between constituents of up to three-spin subsystems, shown in Fig.\ref{fig:fig5}. The oscillatory behaviour of the former has previously been noted in \cite{Nicola2021}, and is being shown here again for completeness in left and middle subfigures in Fig.\ref{fig:fig5}. However, we point out an interesting feature here. Note that, the maxima of MI($a:a+1,a+2$) and odd-numbered minima (first, third,...) of MI($a:a+1,a+3$) occur at approximately the same points of time, whereas the minima of the former and even-numbered minima of the latter approximately occur together. The latter minima also coincide with the maximum values ($\sim 0$) of the TMI (right, Fig.\ref{fig:fig5}), as well as the minima of entanglement entropies in Fig.\ref{fig:fig1} and maxima of level$-k$ sums in Fig.\ref{fig:fig4}. At these points of time, the subsystems in questions transiently become rather trivial in the sense of having low or negligible entanglement and correlations amongst each other. 
\vspace{0.2cm}

More interesting here is the implication of the oscillatory behaviour of TMI (Fig.\ref{fig:fig5}, right). Considering the discussion in section \ref{sec2c} regarding TMI (Eq.\ref{TMI}) and its interpretation as an indicator of scrambling of local information, it is evident that a series of scrambling and then \textit{un}scrambling take place very noticeably between three neighboring spins, and again a time scale of $\sim 1.55-1.56$ seconds exists between consecutive minima or maxima. The unscrambling behaviour attests to the slow relaxation/thermalization of the paramagnetic-to-ferromagnetic quench dynamics. Between a pair of consecutive maximum and minimum, an increasingly negative TMI between spins (ABC) implies, by arguments of \cite{Hosur2016}, increasing difference of (bipartite) mutual information that spins (BC) together have about spin (A) compared to what spin (B) and spin(C) separately have about spin (A) (this is the scrambling of local information about (A) onto the composite (BC)), and then between a pair of consecutive minimum and maximum, this gets reversed and TMI reaches almost zero after every such cycle. This reversal indicates a gradual \textit{recovery} (unscrambling) of mutual information about spin (A) that (B) and (C) separately have from the composite (BC). It would be interesting in future to find out if this dynamics may be modeled sufficiently accurately using the framework of recovery maps in quantum information theory and consequent implications for using this quench dynamics as a platform for quantum informational experiments.

\begin{figure}[tp]
    \centering
    \includegraphics[width=6cm,height=5cm]{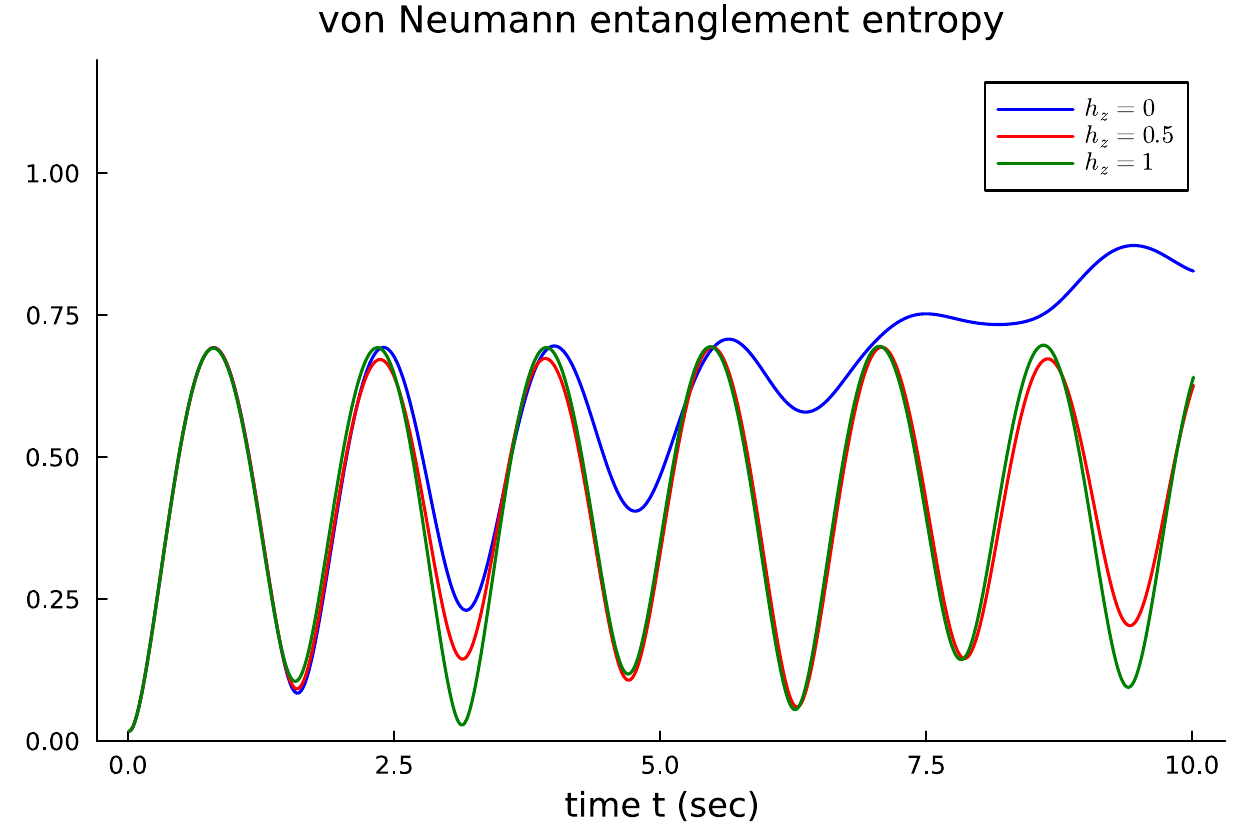}
    \hspace{0.5cm}
    \includegraphics[width=6cm,height=5cm]{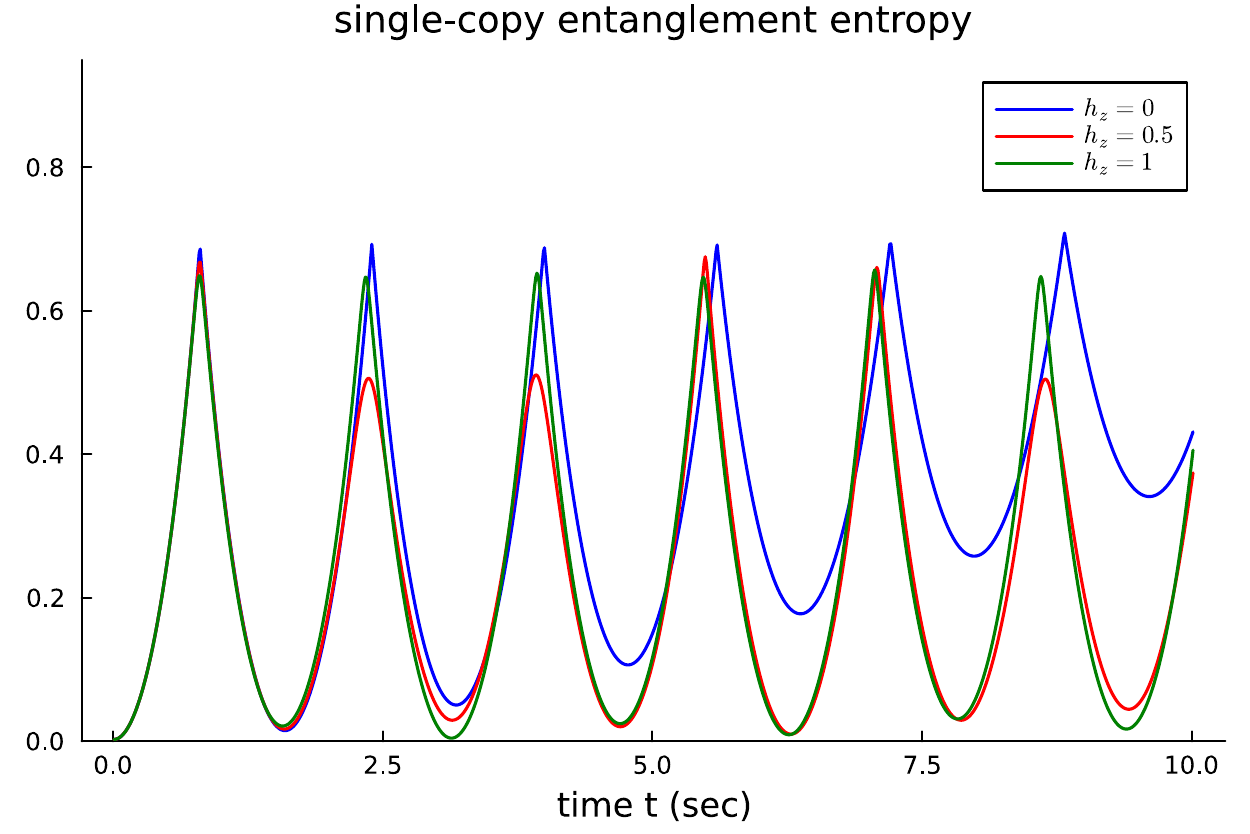}
    \caption{\fontsize{10}{12} \selectfont Half-chain von Neumann (\textbf{left}) and single-copy (\textbf{right}) entanglement entropy at various $h_z$ values for paramagnetic-to-ferromagnetic quenches.}
	\label{fig:fig6}
\end{figure}

\subsubsection{\textbf{Non-integrable vs. integrable quenches}} We next show the dependence of these attributes on the level of non-integrability of the quenching Hamiltonian. All of these features are retained, and in fact strengthened, with increasing non-integrability (higher $h_z$) whereas the integrable point ($h_z\!=\!0$) quickly loses these features and approaches faster towards its equilibrium state (describable by a generalized Gibbs ensemble). A selection of these are shown in Figs.\ref{fig:fig6},\ref{fig:fig7},\ref{fig:fig8},\ref{fig:fig9}. Several remarkable aspects are revealed here :
\vspace{0.2cm}

\begin{figure}[tp]
    \centering
    \includegraphics[width=5cm,height=4cm]{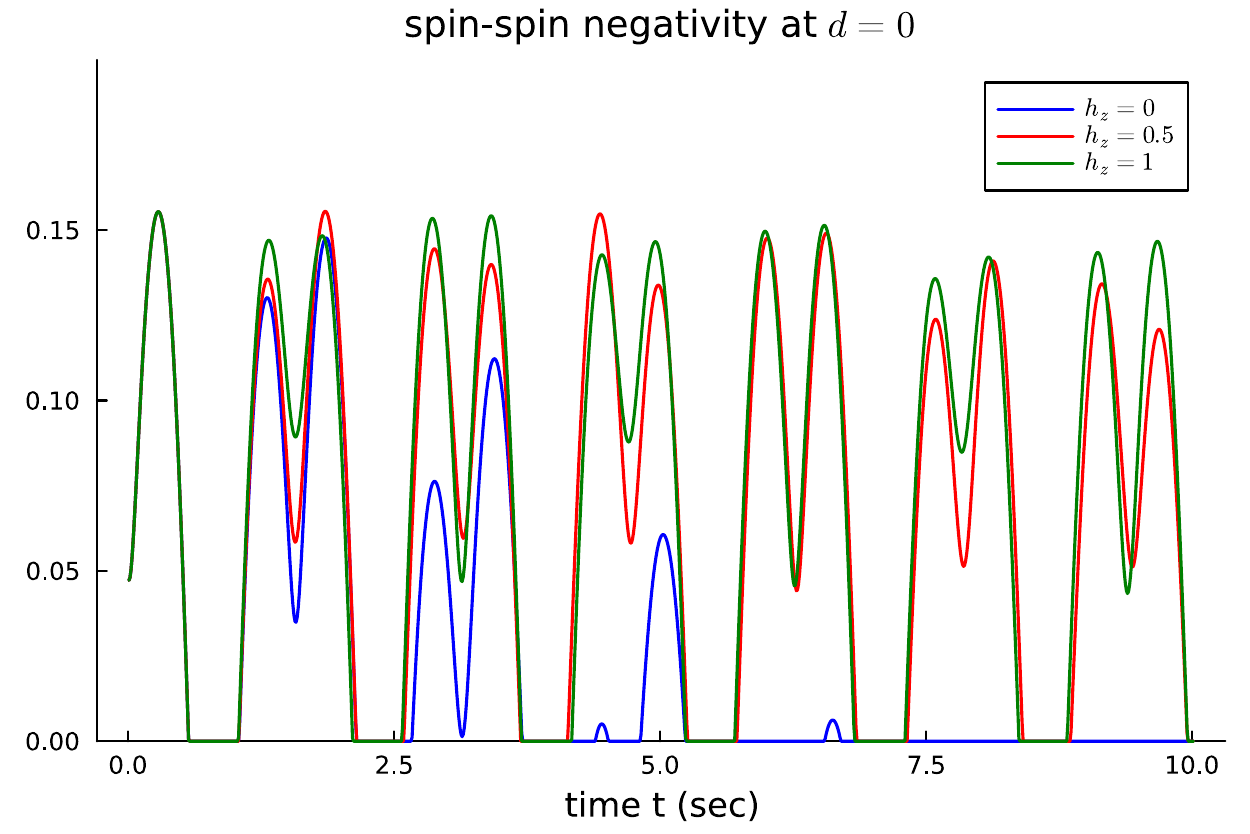}
    \hspace{0.5cm}
    \includegraphics[width=5cm,height=4cm]{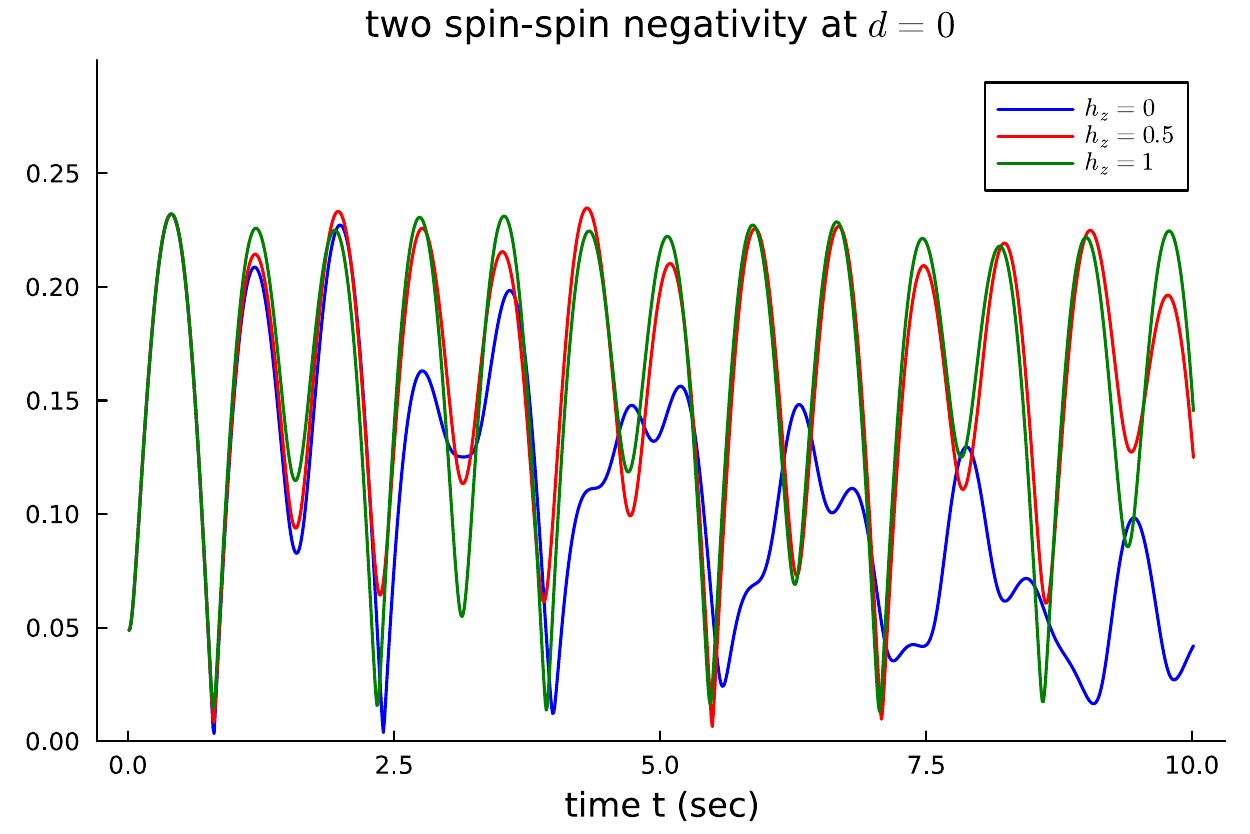}
    \caption{\fontsize{10}{12} \selectfont Negativity of an adjacent spin-spin pair (\textbf{left}) and two-spin with an adjoining third spin (\textbf{right}) at various $h_z$ values for paramagnetic-to-ferromagnetic quenches.}
	\label{fig:fig7}
\end{figure}

\begin{figure}[h]
    \centering
    \includegraphics[width=5cm,height=4cm]{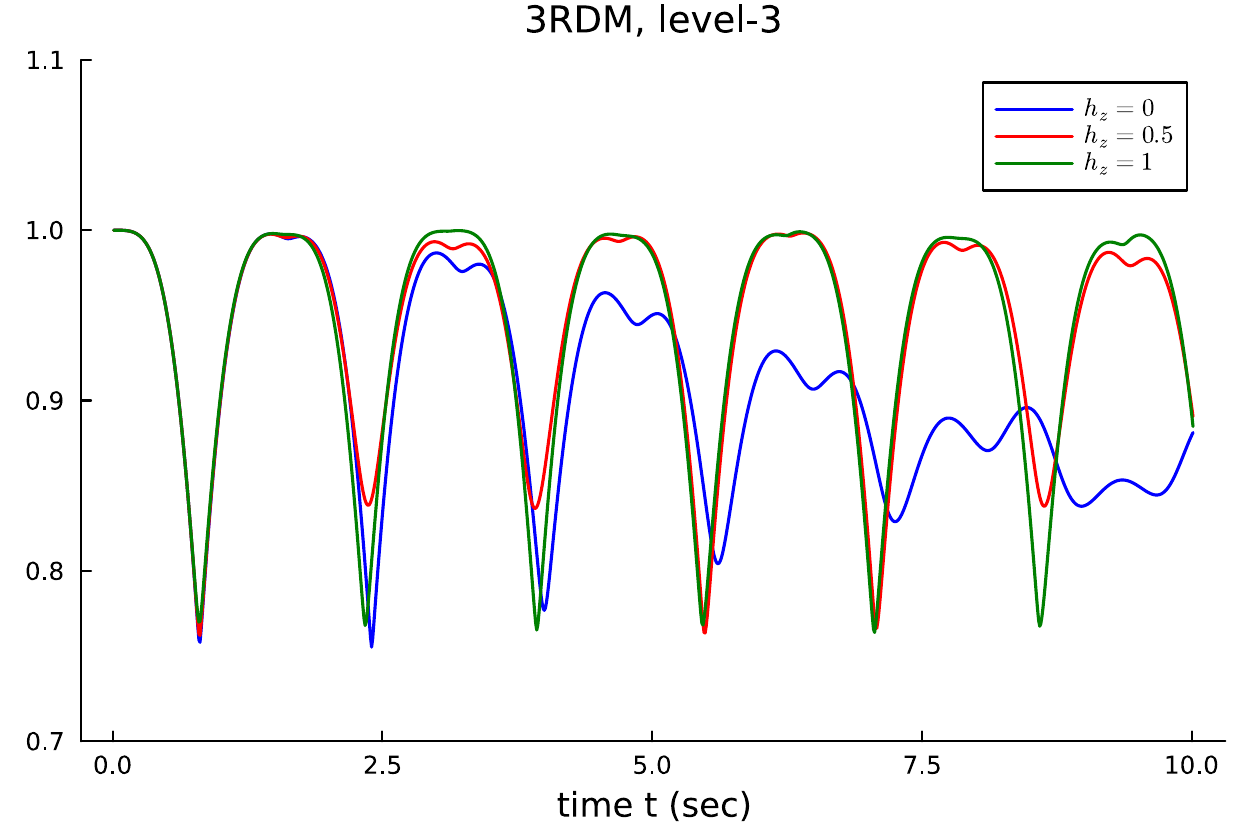}
    \hspace{0.5cm}
    \includegraphics[width=5cm,height=4cm]{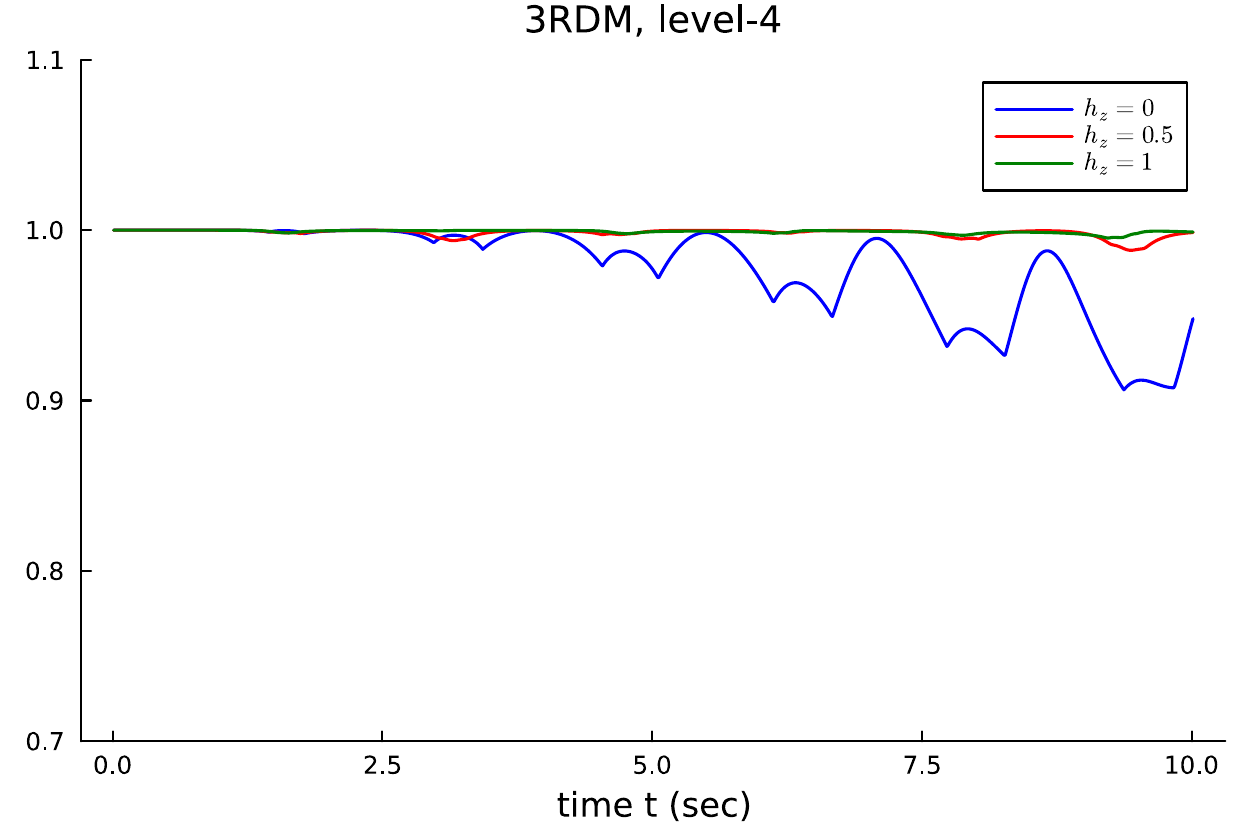}
    \caption{\fontsize{10}{12} \selectfont Level$-3$ (\textbf{left}) and level$-4$ (\textbf{right}) eigenvalue sums of a three-spin subsystem's reduced density matrix (referred to as 3RDM) at various $h_z$ values for paramagnetic-to-ferromagnetic quenches.}
	\label{fig:fig8}
\end{figure}

 (\textit{\textbf{a}}) In all considered quantities, the integrable and non-integrable cases show nearly overlapping behaviour for the first $\sim 2-2.5$ seconds, and it is only after this initial period that features characteristically distinguishing between the integrable and non-integrable cases begin to show up.
 \vspace{0.2cm}
 
 (\textit{\textbf{b}}) As evident from Fig.\ref{fig:fig6} (left), the integrable case ($h_z\!=\!0$) shows a stronger buildup of von Neumann entanglement entropy, with its envelope increasing linearly with time, albeit with accompanying oscillations the strength of which however become milder with time. In contrast, the integrable cases ($h_z\!=\!\{0.5,1\}$) show persistently strong oscillations for a long time and overlapping with each other.
 \vspace{0.2cm}
 
 (\textit{\textbf{c}}) A prominence of non-analytic cusps in single-copy entanglement entropies (Fig.\ref{fig:fig6}, right) is evident in the integrable case, signifying a series of phase transitions in the (ground state space of the) corresponding entanglement Hamiltonian. On the other hand, amongst the two non-integrable cases, the dynamics with $h_z\!=\!1$ shows more number of these cusps compared to that with $h_z\!=\!0.5$. The underlying reason for this set of features is not clear to us at this time but we hope to be able to provide an explanation in future. 
 \vspace{0.2cm}
 
 (\textit{\textbf{d}}) The negativity in both the situations shown in Fig.\ref{fig:fig7} shows qualitatively similar behaviour for both the non-integrable cases but it has a decaying behaviour for the integrable case. Thus, while the integrable quench creates more bipartite pure state entanglement over time (Fig.\ref{fig:fig6}), the opposite is the case with mixed state entanglement between constituents of small subsystems of two or three adjoining spins.
 \vspace{0.2cm}
 
 (\textit{\textbf{e}}) For a three spin subsystem (other subsystems have shown similar behaviour), the corresponding reduced density matrix shows higher mixedness with time as indicated by the dynamics of level-3 and level-4 sums in Fig.\ref{fig:fig8}, where the envelope of these sums decay with time, thus beckoning more participation from the next largest eigenvalue(s) so that the total sum of all eigenvalues remains $=1$. In contrast, the level-4 sums in both the non-integrable cases hovers around the value $1$, meaning only the first three largest eigenvalues show any noticeable dynamics.
 \vspace{0.2cm}

\begin{figure}[bp]
    \centering
    \includegraphics[width=5cm,height=4cm]{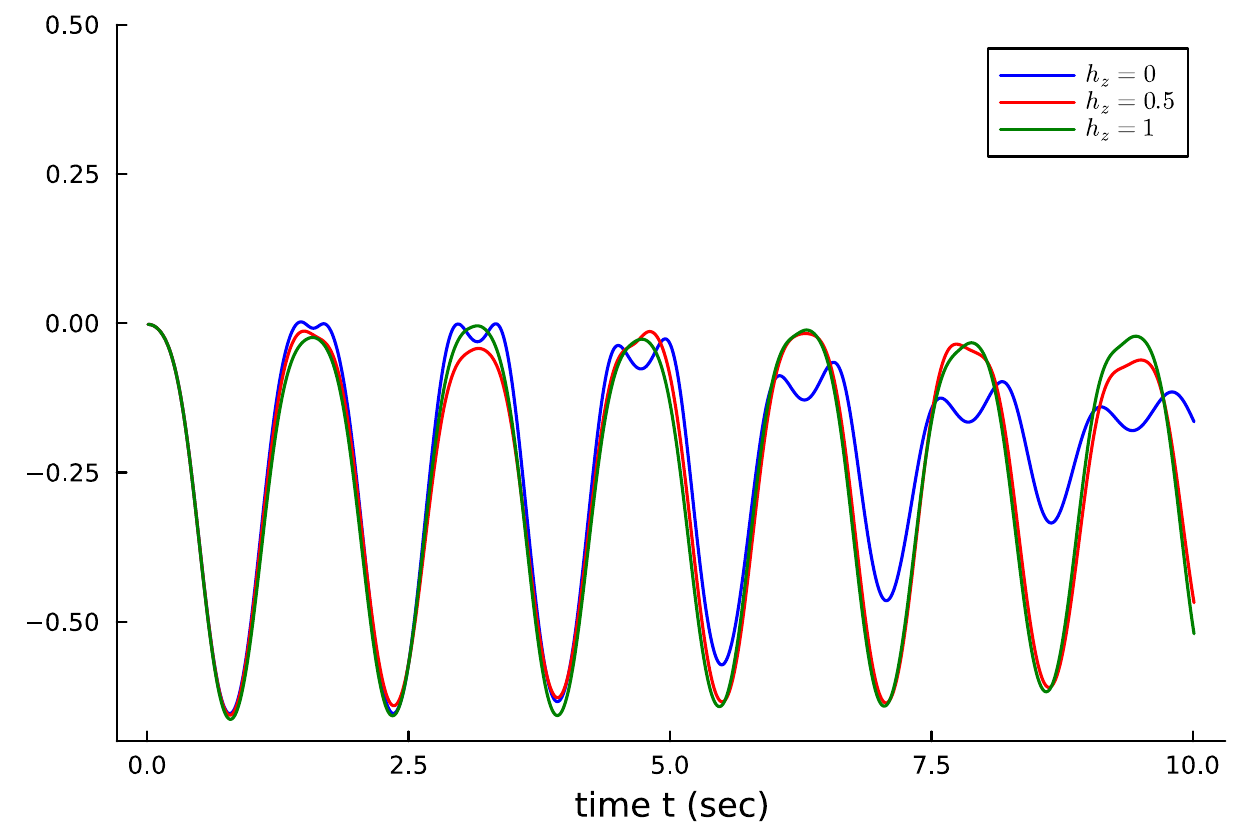}
    \caption{\fontsize{10}{12} \selectfont Tripartite mutual information between three neighboring spins at various $h_z$ values for paramagnetic-to-ferromagnetic quenches.}
	\label{fig:fig9}
\end{figure}
 
 (\textit{\textbf{f}}) Likewise, the tripartite mutual information (TMI) in Fig.\ref{fig:fig9} shows gradually diminishing and less oscillatory behaviour for the integrable case (likely approaching near-zero at very long times) whereas both the non-integrable cases nearly overlap with each other and exhibit persistently oscillatory behaviour. Thus, as far as scrambling locally amongst three neighboring spins is concerned, the non-integrable cases show stronger and more persistent scrambling-unscrambling behaviour likely continuing for long times.

\begin{figure}[h]
    \centering
    \includegraphics[width=5.1cm,height=4cm]{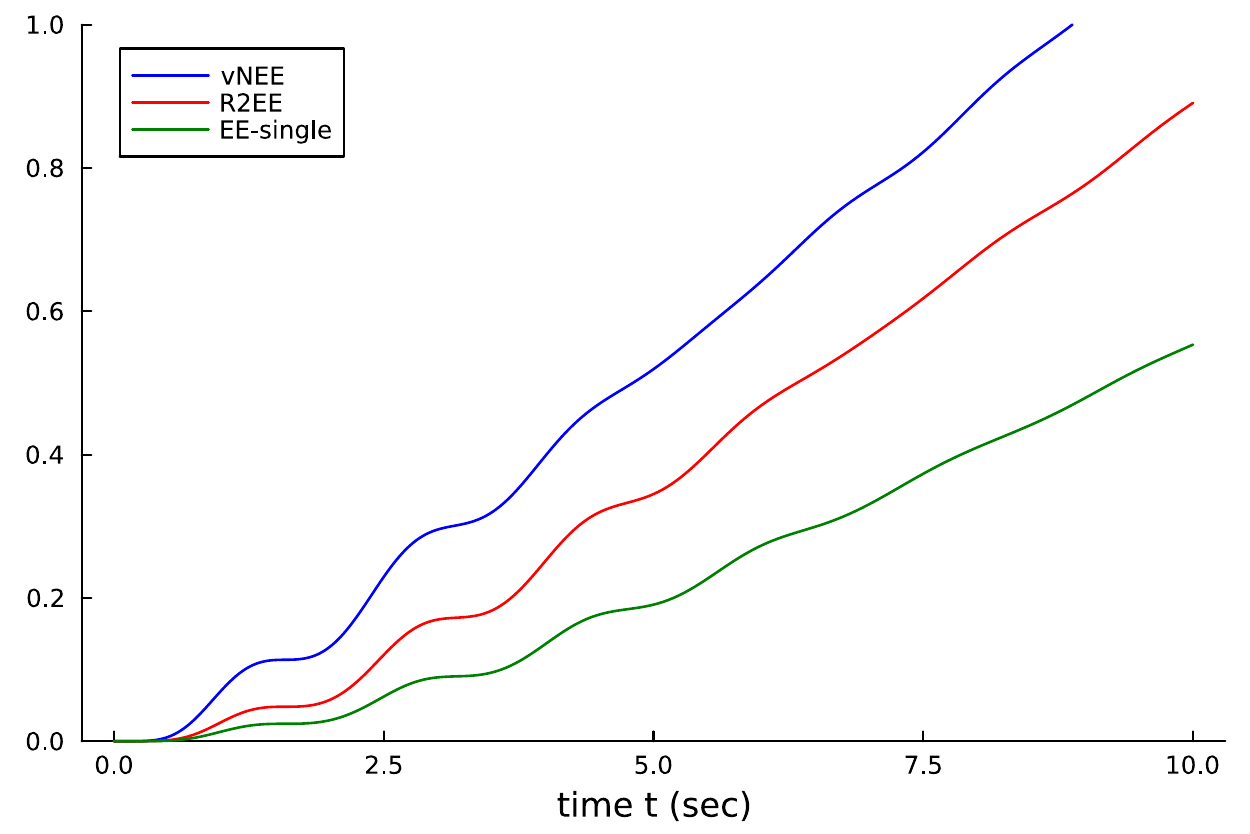}
    %\hspace{0.5cm}
    \includegraphics[width=5.1cm,height=4cm]{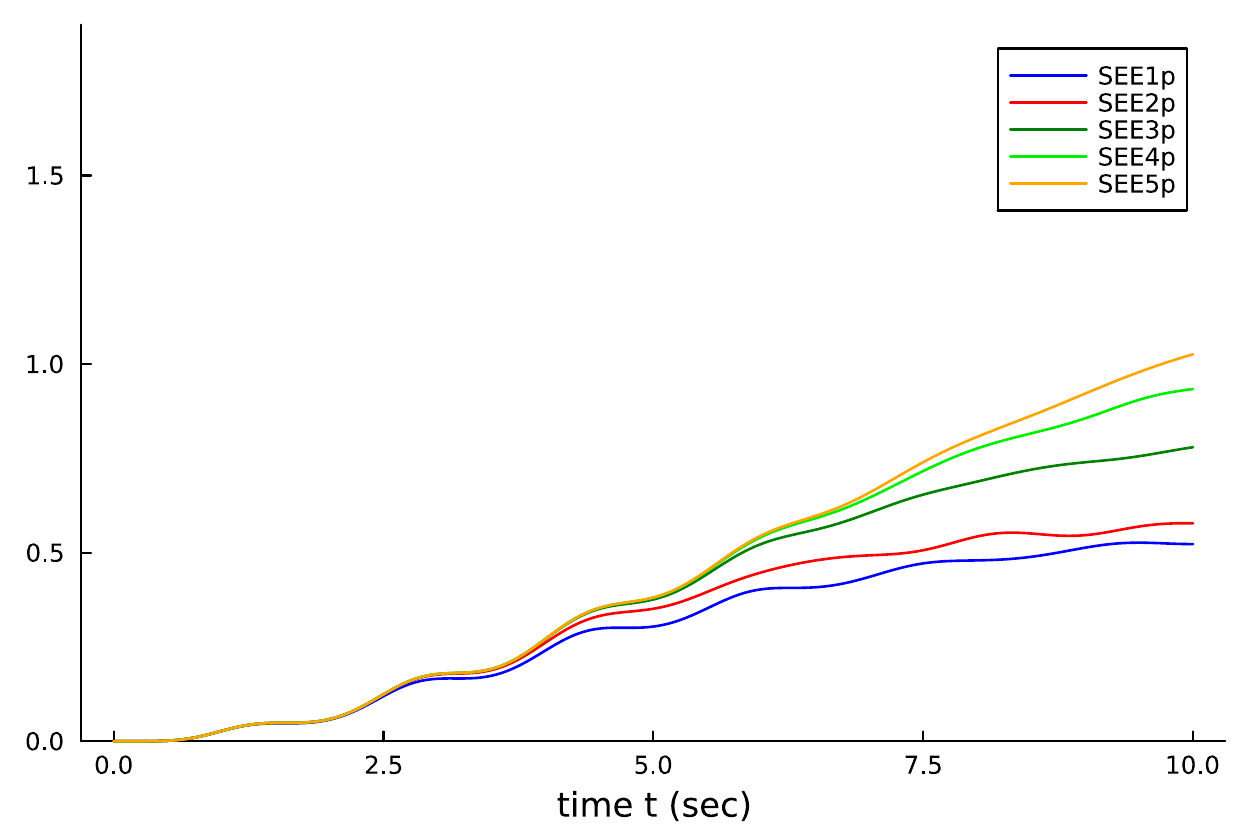}
    %\hspace{0.5cm}
    \includegraphics[width=5.1cm,height=4cm]{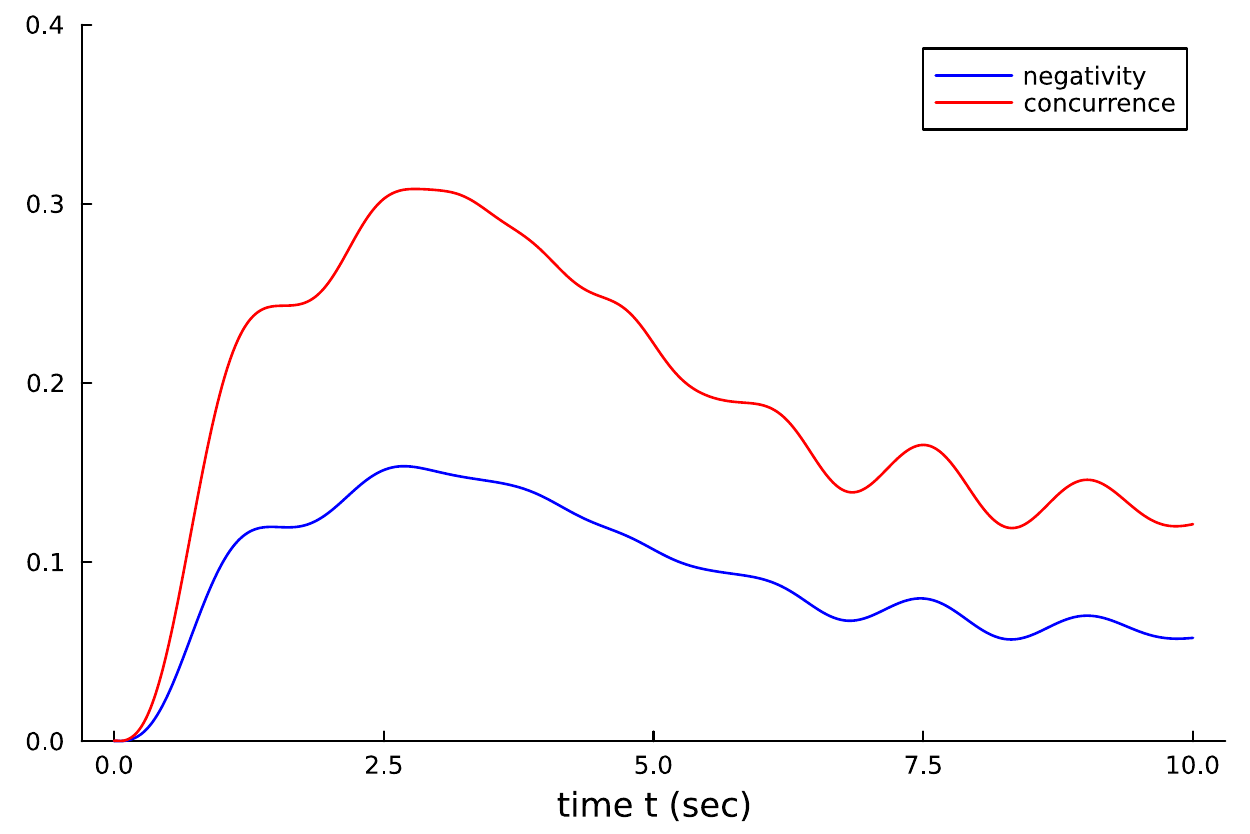}
    %\hspace{0.5cm}
    \includegraphics[width=5.1cm,height=4cm]{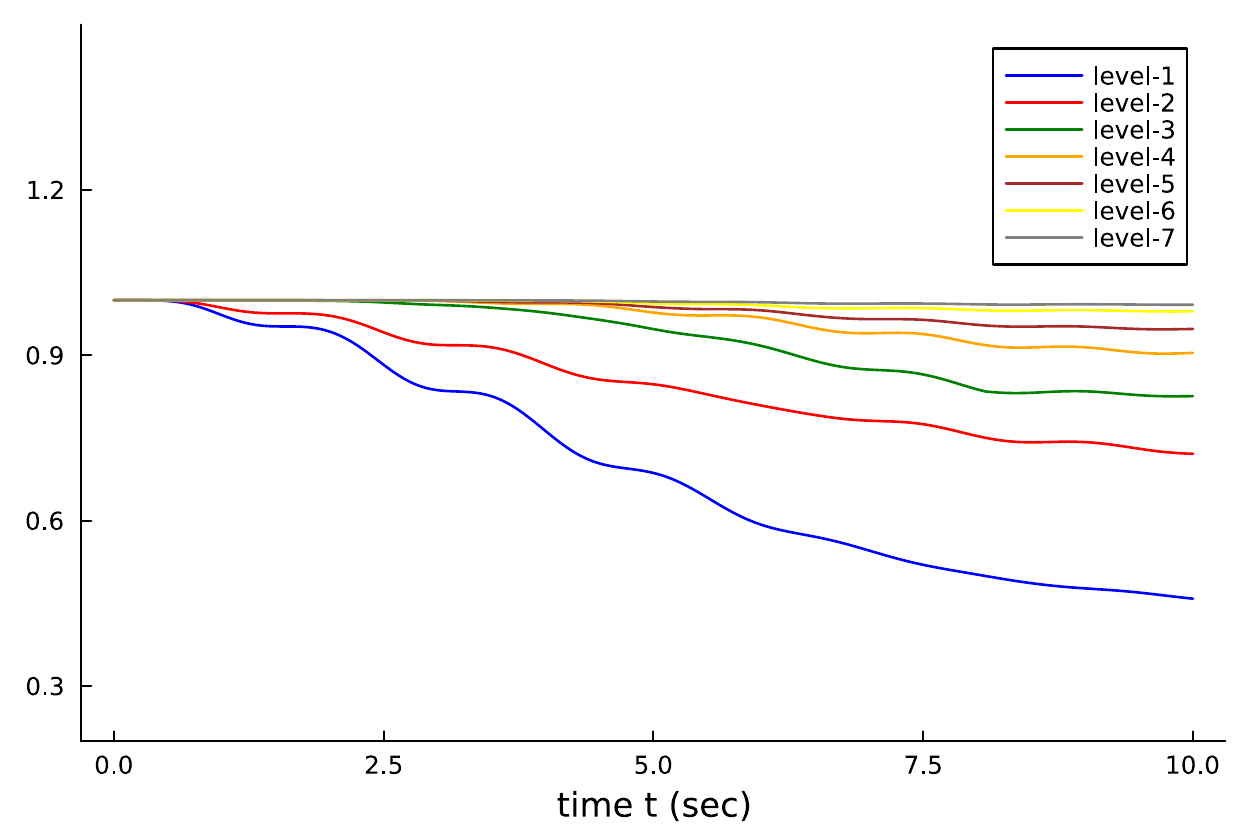}
    %\hspace{0.5cm}
    \includegraphics[width=5.1cm,height=4cm]{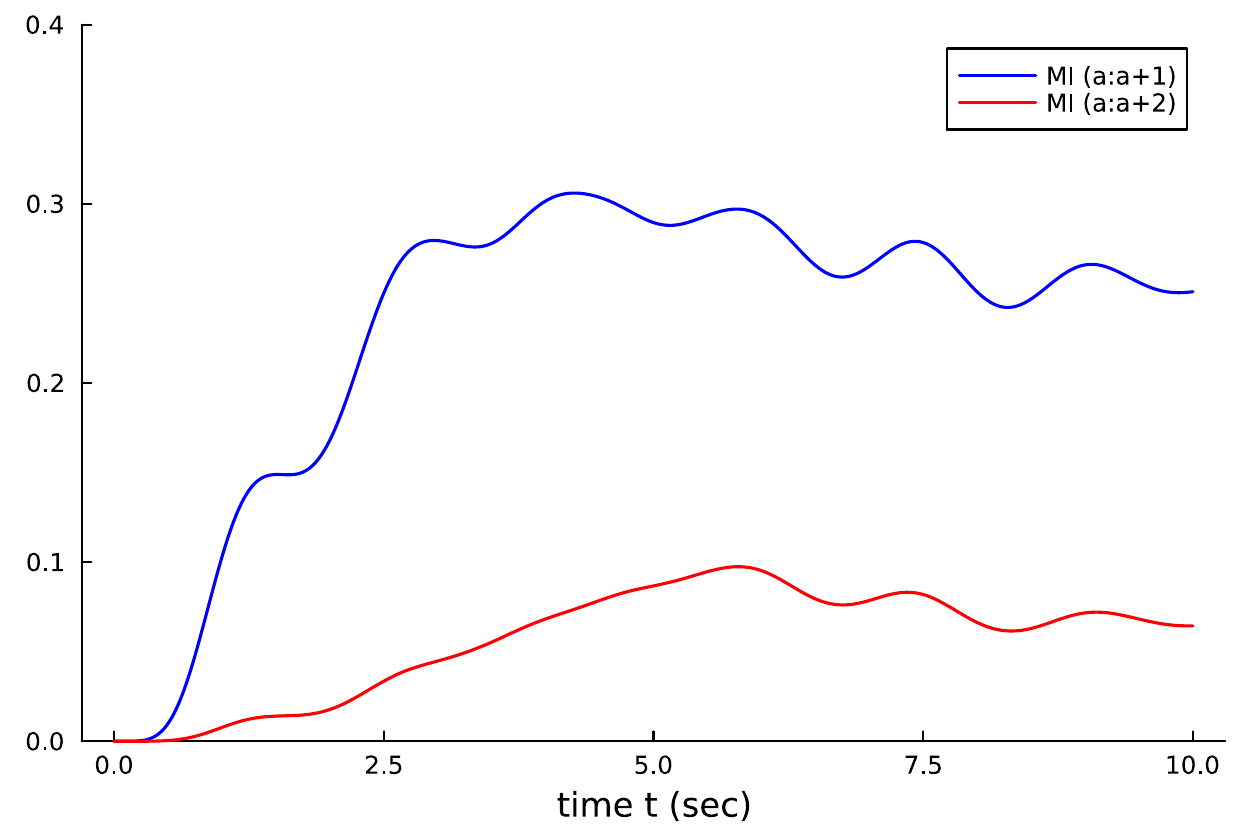}
    %\hspace{0.5cm}
    \includegraphics[width=5.1cm,height=4cm]{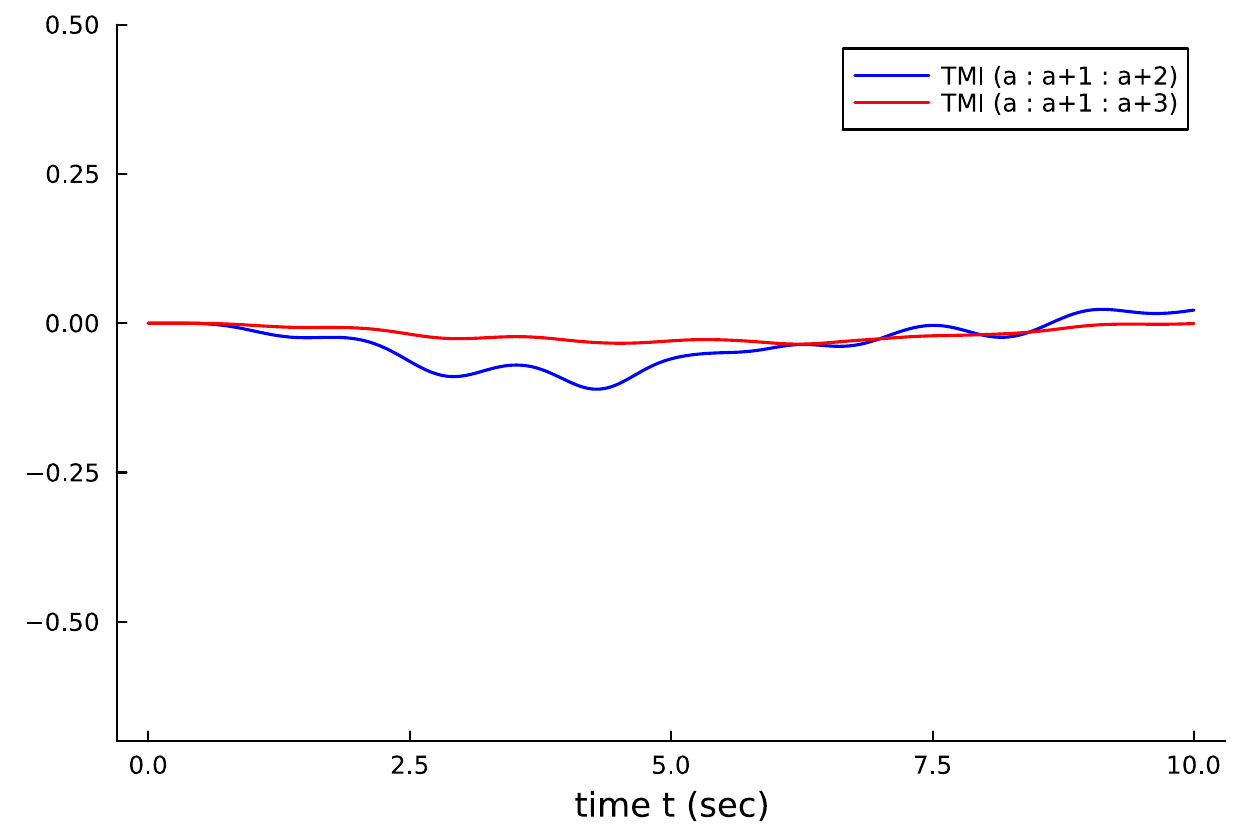}
    \caption{\fontsize{10}{12} \selectfont A selection of features of ferromagnetic to paramagnetic quench $(J,h_x,h_z) \!=\! (1,0.1,0.5) \rightarrow  (0.2,1,0) $. Half-chain entanglement entropies (\textbf{upper row, left}), single-copy entanglement entropies of small subsystems of up to five neighboring spins (\textbf{upper row, center}), mixed state entanglement of a spin-spin pair as measured by negativity and concurrence (\textbf{upper row, right}), level$-k$ sums corresponding to three spin subsystems (\textbf{lower row, left}), bipartite mutual information between a pair of spins, with $a$ denoting the central spin in the spin chain, but this choice is unimportant by translational invariance (\textbf{lower row, center}), and tripartite mutual information amongst three spins (\textbf{lower row, right}).}
	\label{fig:fig10}
\end{figure}

\subsection{Ferromagnetic to paramagnetic quench} We now compare with the results for the opposite quench from a ferromagnetic ground state at $(J,h_x,h_z)\!=\!(1,0.1,0.5) $ to paramagnetic side $(J,h_x,h_z)\!=\!(0.2,1,0) $.  Simulation times $t$ are now in units of $h_x^{-1}$, denoted by (sec) in the figures on x-axis. We show a selection of results in Fig.\ref{fig:fig10}. This quench shows rather featureless dynamics of the quantum informational entities dealt with in this work, and shows a quick buildup of entanglement and approach to equilibration to a putative generalized Gibbs state (since the paramagnetic quench parameters are in the integrable regime), and the picture of entanglement spreading based on freely propagating quasiparticles (\cite{Calabrese2005,Rieger2011}) approximately holds, as for instance evidenced by the almost linearly increasing bipartite pure state entanglement entropies (Fig.\ref{fig:fig10}, upper row, left) \cite{Chiara2006,Bravyi2007,Kim2013}. Dynamical behaviour of spin-spin mixed state entanglement as measured by concurrence and negativity (Fig.\ref{fig:fig10}, upper row, right) and spin-spin mutual information (Fig.\ref{fig:fig10}, bottom row, center) are also quite unremarkable, with an initial increase and then a gradual decay with weak oscillatory fluctuations, and tripartite mutual information amongst three spins (Fig.\ref{fig:fig10}, bottom row, right) is also featureless and in fact hovers close to zero at all simulation times, showing a rather weak local (at the level of these three spins) scrambling of mutual information. The level$-k$ sums of three-spin subsystems (Fig.\ref{fig:fig10}) also show nearly-monotonically decreasing behaviour (other subsystems also showed similar behaviour), signifying Markovianity of the dynamics of these subsystems (assuming the putative connection between monotonicity of majorization relations and Markovianity indicated previously), in agreement with the discussion in \cite{Banerjee2025}, and moreover, more number of three spin subsystem's RDM eigenvalues visibly participate in the dynamics of these level$-k$ sums as time progresses, indicating increasing levels of mixedness of these subsystems over time.

\section{Conclusion}    \label{sec4}

In this work, we have numerically uncovered several fine-grained features of the dynamics of certain quantifiers of quantum entanglement and information for non-perturbative quenches far across the Ising critical point in one dimension. Contrasting behaviour is seen between paramagnetic-to-ferromagnetic quench and its reverse, with several notable salient features already summarized in the introduction Sec.\ref{sec1}. Moreover, comparisons between the integrable regime (zero longitudinal field) and deep into the non-integrable regime (increasingly strong longitudinal field), where the latter case is known to result in slow thermalization dynamics owing to the confinement of excitations, reveals an overlapping of these dynamical features in the very early times, with subsequently the integrable case exhibiting signatures of better mixedness of subsystems' attributes and a quicker approach to equilibration, whereas stronger levels of non-integrability exhibit more resilience in the oscillatory behaviour of entanglement and mutual information quantifiers. 
\vspace{0.2cm}

That the paramagnetic-to-ferromagnetic quench dynamics, even before considering the effect of longitudinal fields resulting in the confinement of excitations, should be "slow" is also expected by recalling that the initial paramagnetic state is rather structureless and essentially unentangled whereas the target (ferro)magnetic state is structured and highly entangled that takes the form of a generalized $N-$body Greenberger–Horne–Zeilinger (GHZ) state. A dynamics starting from the former and targeted to the latter requires substantial rearrangement within the system so that it can approach and eventually settle into the latter state, and this is apparently quite difficult a task for the system to establish quickly and efficiently, leading to slow equilibration dynamics in the far-from-equilibrium regime. Confinement of kink-antikink excitations in the case of non-zero longitudinal field adds another independent factor to the slow dynamics, with the sign of the longitudinal field breaking the up-down $Z_2$ symmetry of the generalized GHZ state, which now requires the system to rearrange itself to go from product state of $|+\rangle \!=\!(|\uparrow\rangle + |\downarrow\rangle)/\sqrt{2} $ at each site to that of either $|\uparrow\rangle$ or $|\downarrow\rangle$ at each site, depending on the sign of the longitudinal field. However, when a non-zero longitudinal field is present, the difficulty of rearrangement of the global state is a secondary reason for slow dynamics, as the rearrangement from a product state of $|\uparrow\rangle$ or $|\downarrow\rangle$ at each site to a product state of $|+\rangle$ or $|-\rangle$ at each site is not as difficult as the aforementioned generalized-GHZ case. Instead, while the far-from-equilibrium slowness of dynamics in the integrable case can be physically reasoned by the state rearrangement argument above, in the non-integrable case the extra slow dynamics must be primarily caused by confinement of excitations which prohibits propagation of excitations across the systems (emanating from one subsystem and propagating to another subsystem far away) \cite{Kormos2016,Robinson2019,Scopa2022,Birnkammer2022}, leading to an inefficient and slow rearrangement of the global Hilbert space, and signatures of significant non-Markovianity (memory-effects and information backflows) in the dynamics of subsystems \cite{Banerjee2025}.
\vspace{0.2cm}

We expect these results to qualitatively hold for quenches across the Ising (and Ising-type) quantum critical points in other more complicated systems including two dimensional systems. As already mentioned in the main text, it would be worthwhile to (\textit{i}) analytically study the phenomenon of unscrambling of local information in terms of recovery maps, as indicated by the periodic revivals of tripartite mutual information amongst three spins, (\textit{ii}) elaborate on the physical significance and perhaps obtain an analytical description of the non-analytic cusps in single-copy entanglement entropy and consequent phase transitions in corresponding entanglement Hamiltonians, (\textit{iii}) investigate the putative general connection between satisfaction or violation of majorization relations and (non-)Markovianity of non-equilibrium dynamics, (\textit{iv}) explore the occurrences of $k-$uniform states for $k>1$ for finite periods of time and the consequent intrinsic potential of dynamical quantum error correcting capabilities of generic quantum many-body systems, and (\textit{v}) analyze the recurrent Page-like dynamics of the entanglement entropies, potential connections with (non-)Markovianity of dynamics, and other signatures of this phenomenon on general quantum information dynamics. Investigating the effects on these features of making weak and projective measurements on one or more subsystems will be worthwhile as well, as well as the dynamics of other important quantum informational attributes such as multipartite entanglement and discord, among others. We hope to report progress on some of these directions in future.
\vspace{0.2cm}

\textbf{Acknowledgments---} Financial support from the Department of Atomic Energy, India is gratefully acknowledged.
\vspace{0.2cm}

\textbf{Data Availability Statement---} All codes for numerical simulations and data obtained from them on which this work is based can be obtained from the author upon reasonable request.

\textbf{Competing Interests---} The author has no competing or conflict of interests.

%\section*{References}

\bibliography{biblio}     

\end{document}